\documentclass[fleqn,usenatbib]{mn2e}

\usepackage{ae,aecompl}

\usepackage{graphicx}	
\usepackage{amsmath}	
\usepackage{amssymb}	

\title[Modelling the temporal behaviour of Cygnus X-1]{A stochastic propagation model to the energy dependent rapid temporal behaviour of Cygnus X-1 as observed by {\it AstroSat} in the hard state}

\author[Bari et al.]{
Bari Maqbool,$^{1}$
M. Sneha Prakash,$^{2}$ \thanks{E-mail: sneha.m@res.christuniversity.in; shivappa.b.gudennavar@christuniversity.in}
R. Misra,$^{1}$
J. S. Yadav,$^{3}$
S. B. Gudennavar,$^{2}$ 
\newauthor
S. G. Bubbly,$^{2}$
A. Rao,$^{4}$
S. Jogadand,$^{5}$
M. K. Patil,$^{5}$
S. Bhattacharyya,$^{3}$
\newauthor
and K. P. Singh$^{6}$
\\\\
$^{1}$Inter-University Centre for Astronomy and Astrophysics, Ganeshkind, Pune-411007, India\\
$^{2}$Department of Physics and Electronics, CHRIST (Deemed to be University), Hosur Road, Bengaluru-560029, India\\
$^{3}$Tata Institute of Fundamental Research, Homi Bhabha Road, Mumbai-400005, India \\
$^{4}$Department of Physics and Astronomy, Faculty of Physical Sciences and Engineering, University of Southampton, \\ Southampton SO17 1BJ, UK 0000-0003-3105-2615 \\
$^{5}$Swami Ramanand Teerth Marathwada University, Nanded-431 606, India \\
$^{6}$Indian Institute of Science Education and Research Mohali, Knowledge City, Sector 81, SAS Nagar, Manauli 140306, India
}

\date{Accepted XXX. Received YYY; in original form ZZZ}

\pubyear{2019}

\begin{document}
\label{firstpage}
\pagerange{\pageref{firstpage}--\pageref{lastpage}}
\maketitle
\begin{abstract}
We report the results from analysis of six observations of Cygnus X-1 by Large Area X-ray Proportional Counters (LAXPC) and Soft X-ray Telescope (SXT) on-board {\it AstroSat}, when the source was in the hard spectral state as revealed by the broad band spectra. The spectra obtained from all the observations can be described by a single temperature Comptonizing region with  disk and reflection components. The event mode data from LAXPC provides unprecedented energy dependent fractional root mean square (rms) and time-lag at different frequencies which we fit with empirical functions. We invoke a fluctuation propagation model for a simple geometry of a truncated disk with a hot inner region. Unlike other propagation models, the hard X-ray emission ($>$ 4 keV) is assumed to be from the hot inner disk by a single temperature thermal Comptonization process. The fluctuations first cause a variation in the temperature of the truncated disk and then the temperature of the inner disk after a frequency dependent time delay. We find that the model can explain the energy dependent rms and time-lag at different frequencies.
\end{abstract}

\section{Introduction}\label{intro}
In the last two decades, there have been extensive studies on the rapid temporal behaviour of X-ray binaries primarily using {\it Rossi X-ray Timing Explorer (RXTE)} data. These studies have considerably broadened the understanding of behaviour of matter in the strong gravity regime of innermost regions of accretion disk. The X-ray variability of these sources are characterized by their power density spectra (PDS) which show broad band continuum noise like features and some times peaked features known as quasi-periodic oscillations (QPOs). The origin of the broad band continuum noise could be due to perturbations which occur throughout the disk and propagate inwards causing the X-ray variation \citep{Lyu97}. Since these variations occurring at different radii are expected to have a multiplicative effect on the accretion rate of inner disk, the scenario provides an explanation for the linear relationship of the rms variability with the flux \citep{Utt01,Gle04}. Additionally, the model also makes predictions on energy dependent temporal properties of the system. In particular, the variation of the fractional rms and time-lag as a function of energy can provide clues to the geometry of the system within this paradigm \citep{Bot99, Mis00, Kot01}. The origin of QPOs is not yet clear as to whether they are arising due to resonances in the disk or precision of inner regions.
\\[6pt]
Recently, there have been efforts to develop simple models based on propagation of fluctuations which attempt to make quantitative predictions for the energy dependent PDS and time-lag \citep{Ing11, Ing12, Ing13, Rap16}. The time-lag between the different energy bands is attributed to the time taken for the fluctuations to propagate, which is taken to be of the order of the viscous time-scales. To  explain the frequency dependency of the observed time-lags, these models generally invoke that photons of different energies arise primarily from the different regions of the disk. This is taken into account by an empirical emissivity index \citep{Kot01} or that there are multiple different Comptonization regions producing hard X-rays \citep{Rap17, Axe18}. If the high energy photons are produced by thermal Comptonization from a region having optical depth of order unity, it is not clear how such multiple different temperature regions may exist and whether the emissivity would be a strong function of radius. A radically different view, where hard X-ray photons arise from an optically thick disk with a rapidly varying temperature profile, has been evoked to explain the frequency dependent time-lags \citep{Mis00}.
\\[6pt]
The behaviour of the propagating fluctuations and their effect on the radiative processes active in the disk need to be understood from first principles by simulating realistic accretion disk models. Magnetohydrodynamic simulations for fluctuations propagating in standard disk have proved the basic principles of the model \citep{Cow14, Hog16}. However, one interesting feature of these simulations is that the propagation time-scales can be frequency dependent and are not necessarily the same as the viscous ones. Using simple time dependent standard disk equations, \citet{Ahm18} also showed that this is the case and quantify the frequency dependent behaviours. However, these results are for standard outer disks extending to the last stable orbit and it is not clear how the results would change if the standard outer disk is truncated at some distance from the black hole and has a hot inner flow \citep{Yua14}. Nevertheless, it is important to consider the possibility of a single radiative region with frequency dependent propagation time-scales and to see if such a model can also explain the observations.
\\[6pt]
These theoretical concepts have to be tested by comparing their predictions with the observational data from X-ray binaries. The data should, not only provide detailed timing information but also simultaneous wide band spectral coverage to constrain the spectral parameters. Such an opportunity has been provided by the Indian multi-wavelength observatory {\it AstroSat} \citep{Agr06, Sin14}.
The LAXPC \citep{Yad16a, Agr17} on-board  {\it AstroSat} has a larger effective area at energies above 30 keV as compared to its predecessor {\it RXTE}. LAXPC can measure the arrival time of each photon accurately to about 10 $\mu$s. It always provides event mode data, which allows for detailed energy dependent study of rapid temporal behaviour \citep{Yad16a}. Preliminary analysis of LAXPC data of Cygnus X-1 has already demonstrated these capabilities \citep{Mis17}. Another important advantage is the SXT \citep{Sin16, Sin17} on-board {\it AstroSat} which provides simultaneous soft X-ray spectral coverage. Thus, {\it AstroSat's} ability to provide wide band spectral coverage along with rapid variability information, makes it an ideal observatory to test stochastic propagation models for X-ray binaries.
\\[6pt]
In this work, we report the results from our analysis of six observations of Cygnus X-1 by {\it AstroSat's} LAXPC and SXT. The results for one of the observations by LAXPC have been reported earlier \citep{Mis17} when Cygnus X-1 was in its hard state. First, we attempt to empirically quantify the energy dependent temporal behaviour of the source. Then, we interpret the results in terms of a propagation model wherein the high energy emission is assumed to be emitted from a single zone.\\[6pt]
A description of the {\it AstroSat} observations and the data analysis
is given in the next section. This is followed by the presentation of the LAXPC spectral and timing results in \S 2.1 and \S 2.2 respectively, where the energy dependent rms and time-lag for different frequencies are fitted using empirical functions. The parameters of these empirical functions provide a model independent representation of the data. \S 2.3 describes the  propagation model used in this work and the details of how the rms and time-lag are computed within the framework, while in \S 2.4 the model is tested against the data. Since the propagation model used in this work differs from earlier ones, \S 2.5 provides a comparison of this model with earlier ones and highlights the advantages and limitations of the model. Finally in \S 3, we discuss the results.
\section{Data and Analysis}
\subsection{Spectral analysis}
\begin{figure*}
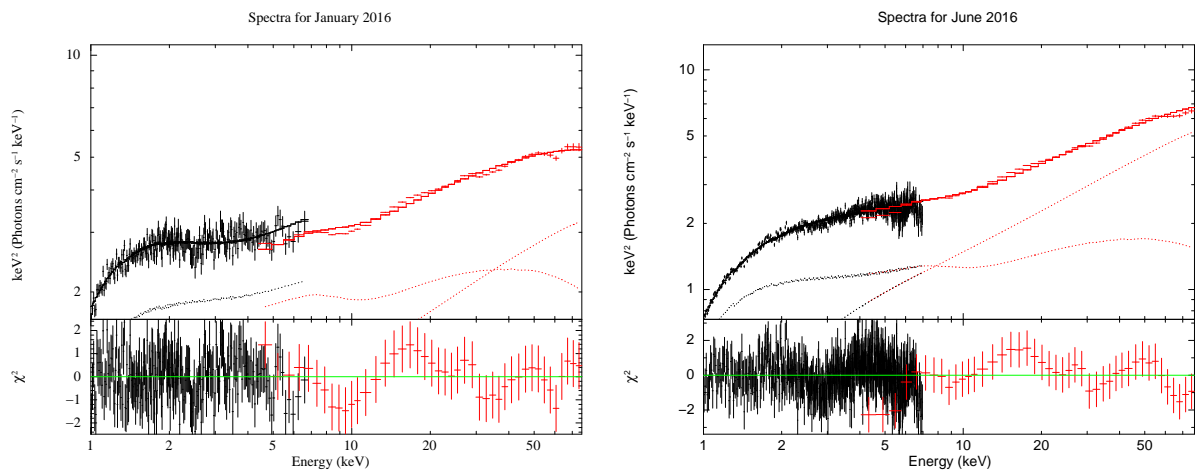

\centering
\includegraphics[width=0.35\textwidth,angle=-90]{janspec.eps}
\includegraphics[width=0.35\textwidth,angle=-90]{junespec.eps}
\caption{Spectra and residuals for January 2016 (Obs ID: 9000000258)(Left panel) and June 2016 (Obs ID: 9000000476)(Right panel).}
\label{spec}
\end{figure*}

We have analyzed six observations of Cygnus X-1 by LAXPC and SXT on-board {\it AstroSat} during 2016 as listed in Table \ref{obstab}. Of these, the LAXPC analysis of January 2016 data (Obs ID: 9000000258) has been reported earlier by \citet{Mis17}. The Level 1 PC mode data of SXT was processed through a SXT pipeline software\footnote{http://astrosat-ssc.iucaa.in/} to produce Level 2 event files. The SXT event merger tool$^1$ was used to create a merged event file from all the clean events files. The HEAsoft tool XSELECT was then used to extract the spectra using source regions within an annulus of inner radius of 1\arcmin and outer radius of 12\arcmin. This was done to account for the pile up effect in the charged coupled device (CCD) due to the high flux of the source. 
\\[6pt]
Analysis of LAXPC and SXT data were carried out using LAXPC software$^1$ and XSELECT respectively. The LAXPC software generates a standard GTI file which removes Earth occultation and the South Atlantic Anomaly (SAA) passes. We have verified that all the three proportional counters (LAXPC 10, 20 and 30) were on by visually examining each lightcurve at 10 second time bins. For SXT data, the images were visually checked to ensure that the drift corrections were applied and a single point source was seen. SXT images of few orbits from January and July which showed multiple points indicating multiple sources were discarded.
\\[6pt]
LAXPC spectra were fitted using XSPEC 12.9.0 in the energy range 4\textendash80 keV. The response and background files used for the fitting were obtained from LAXPC software \citep{Ant17}. The uncertainties in the response calibration require a systematic error of 3\% to be added to the analysis. The background uncertainty of 3\% was also taken into account. We considered the SXT spectra only in the energy range 1-7 keV, since for lower energies there are uncertainties in the effective area and response. Off-axis auxiliary response file (ARF) ``sxt$\_$pc$\_$excl01$\_$v03.arf'' and response matrix file (RMF) ``sxt$\_$pc$\_$mat$\_$g0to12.rmf'', appropriate for the source location on the CCD were obtained using sxtmkarf$^1$ tool. SXT observations of the blank sky were used to extract the background spectrum. A 3\% systematic error and an offset gain correction was applied to the SXT data. The relative normalization was allowed to vary between the LAXPC and SXT data.
\\[6pt]
The spectral analysis was performed using data from LAXPC 20 and SXT following the procedure adopted by \citet{Mis17}. A relatively simple model with the thermal Comptonization ({\it nthcomp}) \citep{Zyc99} as the primary component, a reflection component({\it reflionx}), disk emission ({\it diskbb}) and an interstellar absorption ({\it tbabs}) was used. The disk emission was taken to be the seed photons for the Comptonization. The electron temperature of the Comptonizing medium was fixed to 70 keV. The representative spectral fits are shown in Figure \ref{spec} and the best fit parameters are listed in Table \ref{tablepar1}. The residuals seen in Figure \ref{spec} may be due to uncertainties in the response and background.
\\[6pt]
The high energy spectral indices ($\Gamma \sim 1.5$) found are harder than the usually observed hard state spectral index ($\Gamma \sim 1.6$) of Cygnus X-1 \citep{Ibr05}. The interstellar hydrogen absorption column density ($N_{H}$ $\sim$$3~\times 10^{21}$ cm$^{-2}$), was found to be similar to that obtained by earlier studies \citep{Xia11, Tom14}. While the temperature and normalization of the disk emission are $\sim1$ keV and $\sim50$ respectively for five of the data sets, the January data reveals a cooler disk at $\sim 0.5$ keV and a larger normalization of $\sim 2000$.
\\[6pt]
The large systematics ($3$\%) used in the fitting of SXT and LAXPC spectra do not allow for more sophisticated spectral models and the spectral parameters obtained here are approximate. For example, use of a different reflection model like {\it ireflect} changes the spectral index by $\sim 0.1$ and the inner disk temperature by $\sim 0.1$ keV. Moreover the degeneracy between the inner disk temperature and normalization (i.e. a smaller inferred temperature may lead to a large normalization estimate) is inherent in these spectral analysis. We discuss the impact of these uncertainties later when we describe the model fits to the timing analysis. Nevertheless, we note that, the spectra for all five observations are similar, except for an interesting difference with regard to January data (9000000258). Note that the LAXPC spectra (i.e. in the 4\textendash 80 keV range) are similar for all the observations and the difference is observed primarily in the low energy SXT spectra.
\begin{table*}
\centering
\caption{Observations of Cygnus X-1}
\begin{center}
\begin{tabular}{l c c c c c}
\hline 
\hline
Obs ID & Start Time & End Time	& Exposure Time (ks)\\
 & Date & Date & LAXPC$~~~$SXT \\
\hline 
9000000258 & 19:57:09 & 16:06:47 &  32$~~~~~~~$11 \\
	   & 08-01-2016 & 09-01-2016 &  \\
9000000436  & 20:23:25 & 02:27:01 & 33$~~~~~~~$21 \\
	    & 29-04-2016 & 01-05-2016 &	\\
9000000476  & 10:41:54 & 02:46:43 & 28$~~~~~~~$16\\
	    & 01-06-2016 & 02-06-2016 & \\
9000000524 & 00:01:44 & 18:11:21 & 26$~~~~~~~$14 \\
	   & 01-07-2016 & 01-07-2016 & \\
9000000572  & 01:19:30 & 18:28:43 & 20$~~~~~~~$11\\
	   & 03-08-2016 & 03-08-2016 & \\
9000000722  & 16:49:10 & 06:53:27 & 14$~~~~~~~$8\\
	    & 09-10-2016 & 10-10-2016 & \\
\hline
\end{tabular}
\end{center}
\label{obstab}
\end{table*}
\begin{table*}
\centering
\caption{Spectral fit parameters}
\begin{center}
\begin{tabular}{l l l l l l l l}
\hline 
\hline
 & & &  Observation IDs & & & \\ 
\hline
Model Parameters & Description/ Unit & 9000000258 & 9000000436 & 9000000476 & 9000000524 & 9000000572 & 9000000722 \\
\hline 
\\
$N_{H}$ & Interstellar absorption  & 0.25$^{+0.05}_{-0.05}$ &  0.27$^{+0.04}_{-0.04}$  & 0.35$^{+0.03}_{-0.03}$ & 0.32$^{+0.05}_{-0.03}$ & 0.36$^{+0.03}_{-0.03}$ & 0.35$^{+0.03}_{-0.03}$\\
\\
& $10^{22} cm^{-2}$  & &   &  &  &  & \\
\\
$\Gamma$  &Asymptotic power-law& 1.56$^{+0.07}_{-0.07}$ & 1.53$^{+0.02}_{-0.02}$ & 1.48$^{+0.02}_{-0.02}$ & 1.56$^{+0.01}_{-0.01}$ & 1.46$^{+0.02}_{-0.02}$ & 1.50$^{+0.02}_{-0.03}$\\
  &photon index&  &  &  &  &  & \\
\\
$kT_{e}$ & Temperature at inner  & 0.46$^{+0.09}_{-0.09}$  & 1.06$^{+0.04}_{-0.04}$ & 1.01$^{+0.10}_{-0.08}$ & 1.02$^{+0.08}_{-0.12}$ & 1.06$^{+0.11}_{-0.09}$ & 1.02$^{+0.08}_{-0.10}$\\
 &disk radius&  &  &  &  &  & \\
\\
 & keV &  &  & &  &  & \\
\\
$N_{comp}$  &Normalization factor& 0.36$^{+0.30}_{-0.18}$ & 0.20$^{+0.04}_{-0.03}$ & 0.24$^{+0.05}_{-0.05}$ & 0.85$^{+0.03}_{-0.03}$ & 0.21$^{+0.05}_{-0.03}$ & 0.22$^{+0.05}_{-0.04}$\\
\\
$N_{Ref}$ &Normalization factor & 3.35$^{+0.9}_{-1.1}$  & 0.72$^{+0.20}_{-0.19}$ & 2.05$^{+0.28}_{-0.28}$ & $< 6.0$ & 2.32$^{+0.25}_{-0.28}$ & 2.07$^{+0.26}_{-0.30}$\\
\\
 & $10^{-6}$ &   &  &  &  & & \\
\\
\\
$N_{disk}$  &Normalization factor& 2390$^{+4550}_{-1430}$& 67$^{+12}_{-12}$ & 52$^{+23}_{-19}$ & 57$^{+23}_{-12}$ & 33$^{+22}_{-12}$ & 51$^{+25}_{-17}$\\
\\
$\chi^2$/dof && 390/430  & 988/643 & 668/643 & 672/606 &  694/643 & 731/642\\
\hline
\end{tabular}
\end{center}
\label{tablepar1}
\end{table*}
\subsection{Timing analysis}
\begin{figure*}
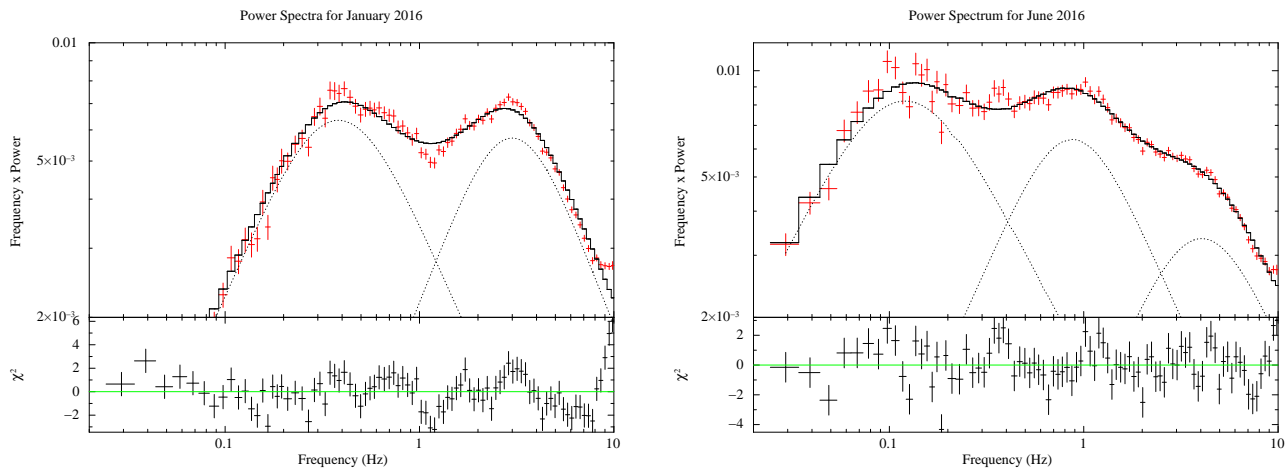

\centering
\includegraphics[width=0.35\textwidth,angle=-90]{janpds.eps}
\includegraphics[width=0.35\textwidth, angle=-90]{junepds.eps}

\caption{The frequency times the PDS of Cygnus X-1 for photons in the 4\textendash 80 keV energy band. For January 2016, the spectra has been fitted with two Lorentzian components (Left panel) and for June 2016, the spectra has been fitted with three Lorentzian components (Right panel). The spectra for other observations are similar to that of June 2016.}
\label{powspec}
\end{figure*}

\begin{figure*}
\centering
\includegraphics[width=0.34\textwidth,angle=-90]{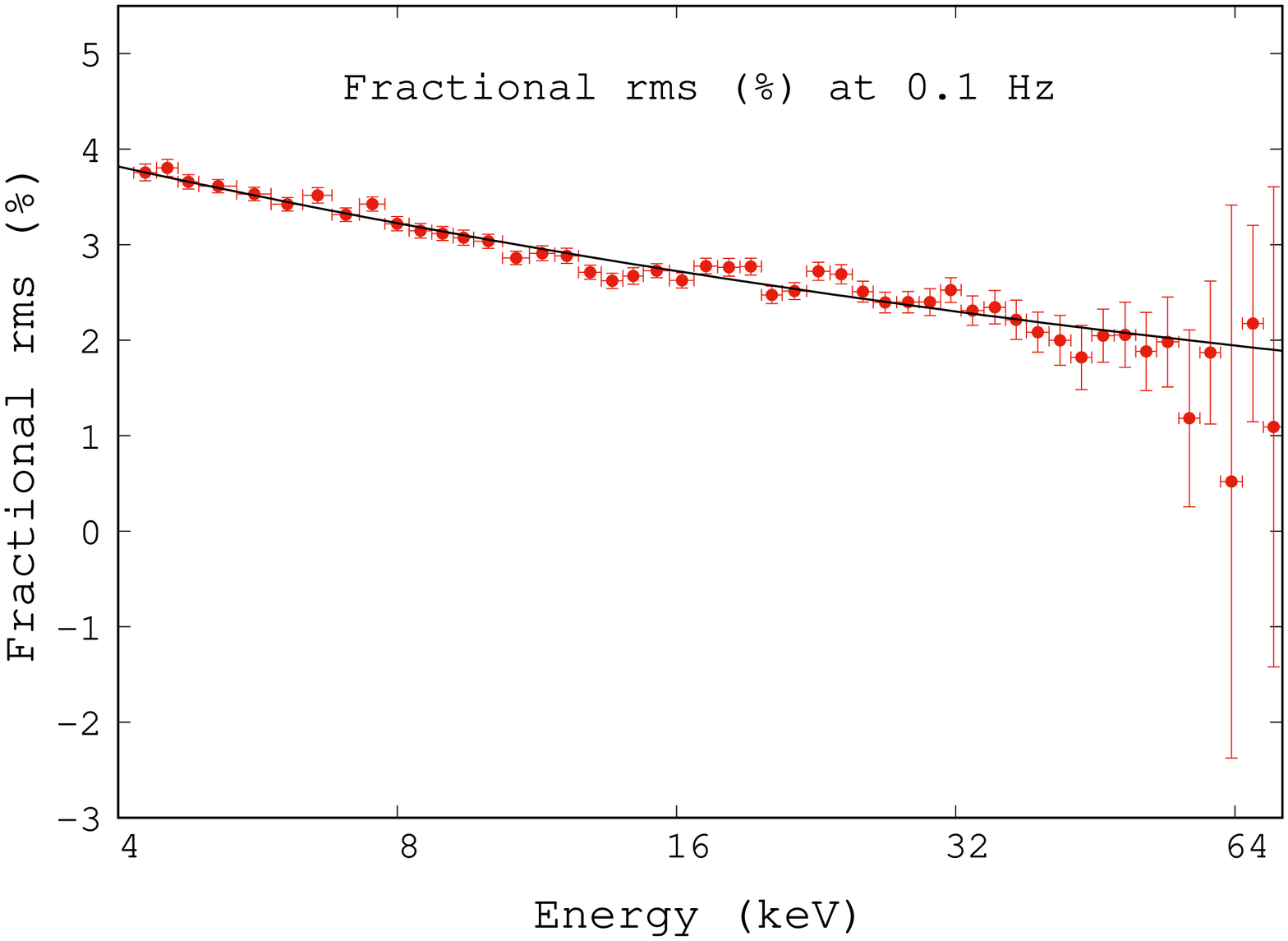}
\includegraphics[width=0.34\textwidth,angle=-90]{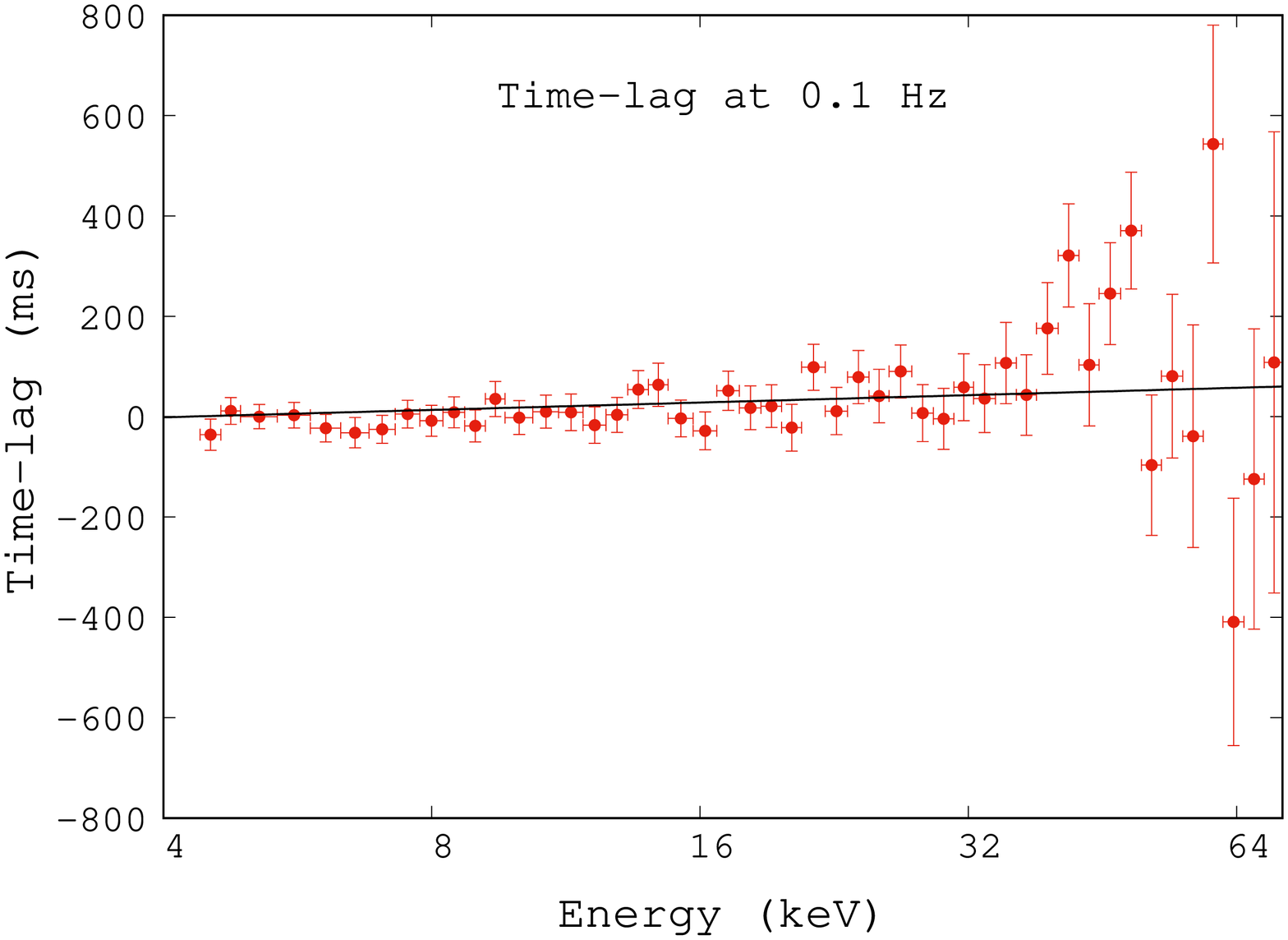}
\includegraphics[width=0.34\textwidth,angle=-90]{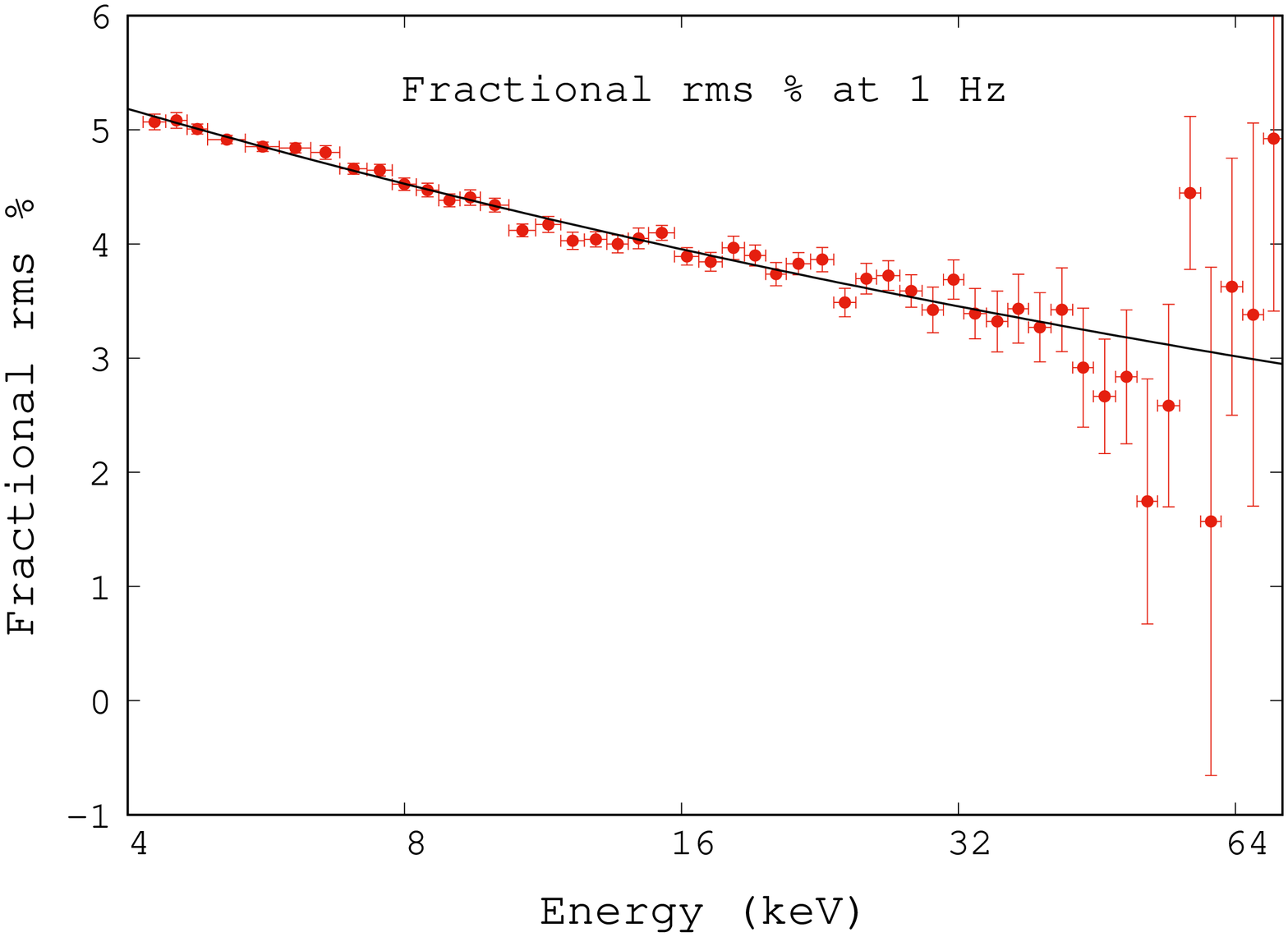}
\includegraphics[width=0.34\textwidth,angle=-90]{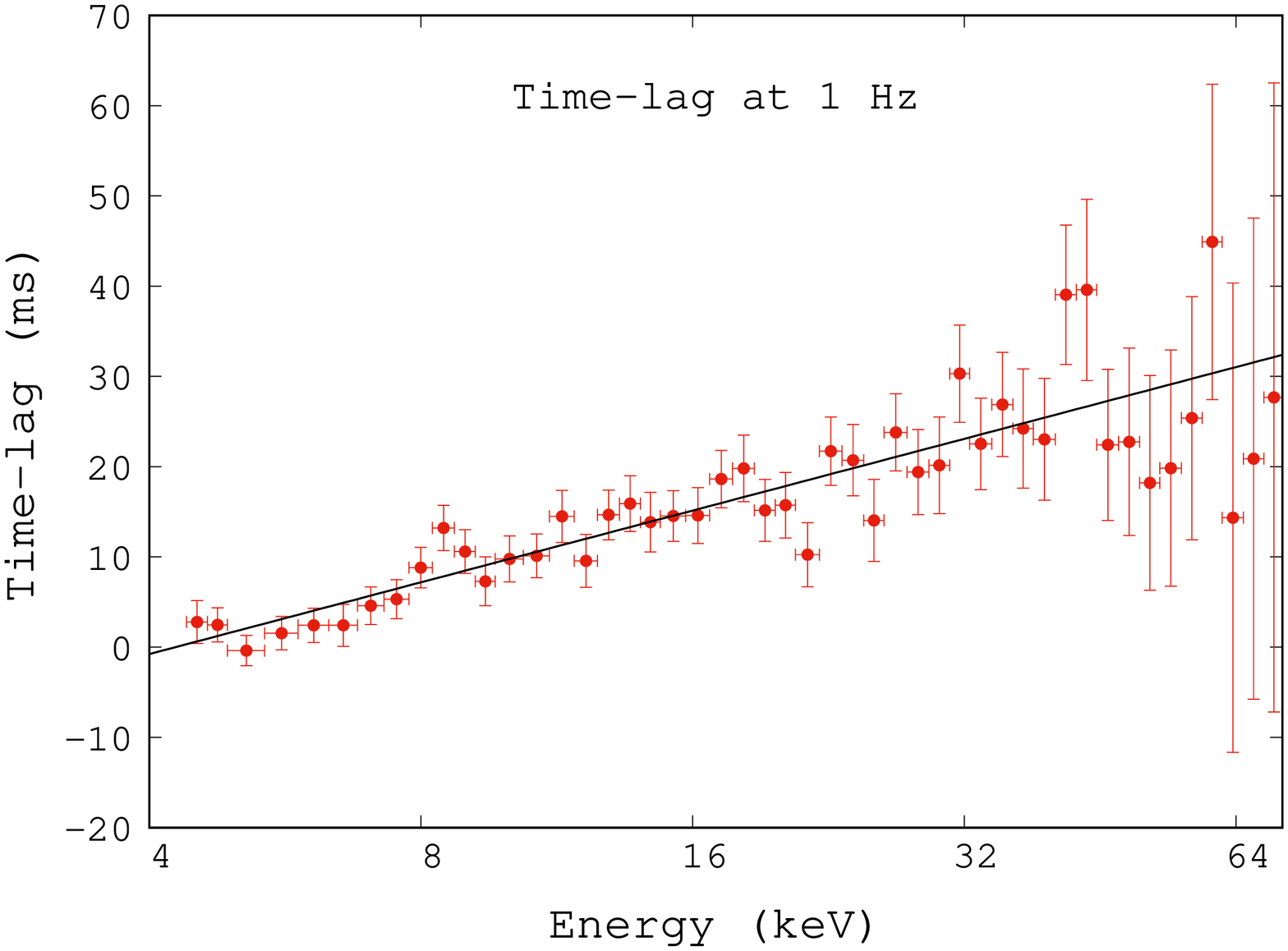}
\includegraphics[width=0.34\textwidth,angle=-90]{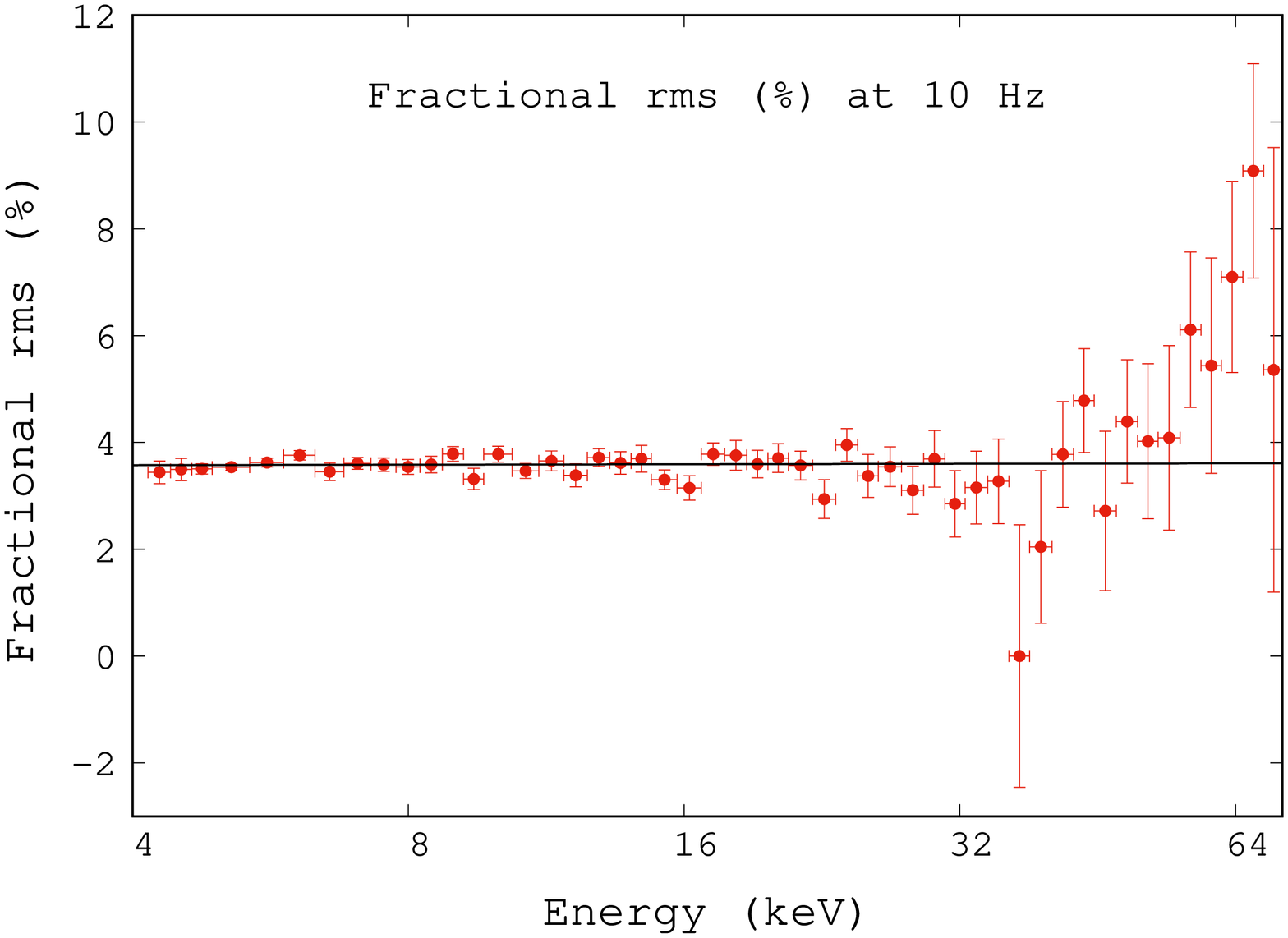}
\includegraphics[width=0.34\textwidth,angle=-90]{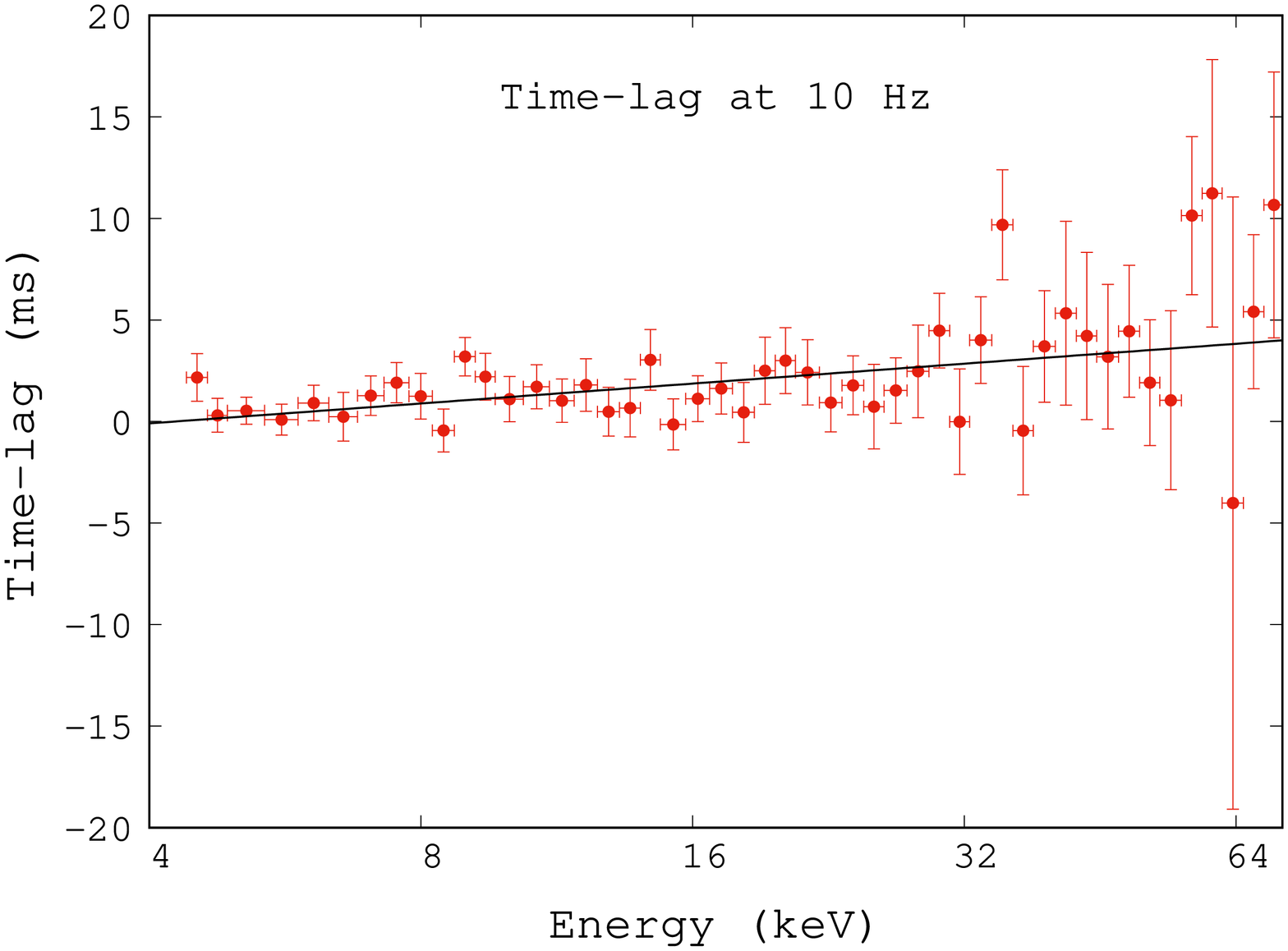}

\caption{Fractional rms and time-lag for three different frequencies 0.1 Hz (top panel), 1 Hz (middle panel) and 10 Hz (bottom panel) for January 2016.}
\label{janrmslag}
\end{figure*}

\begin{figure*}
\centering
\includegraphics[width=0.34\textwidth,angle=-90]{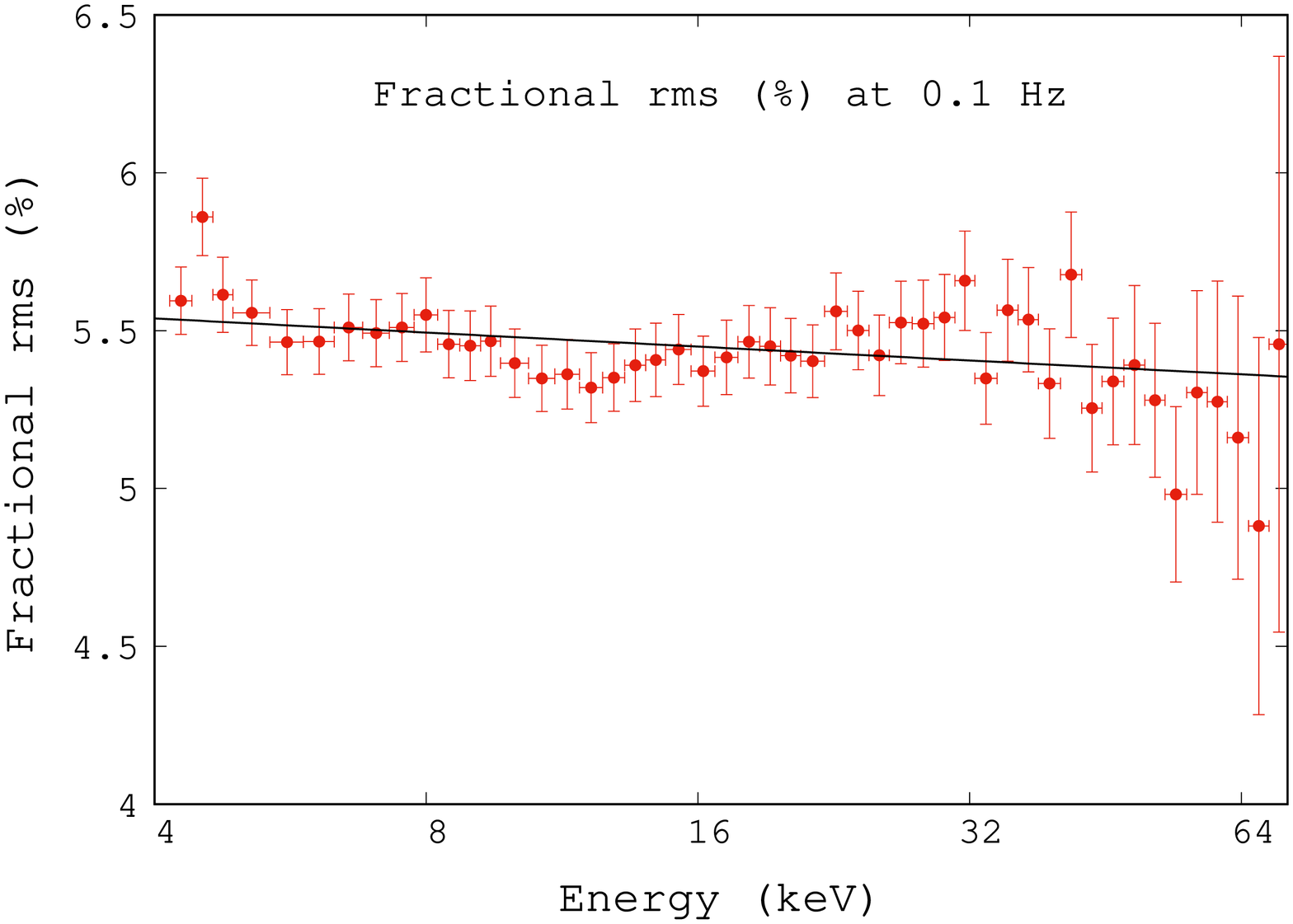}
\includegraphics[width=0.34\textwidth,angle=-90]{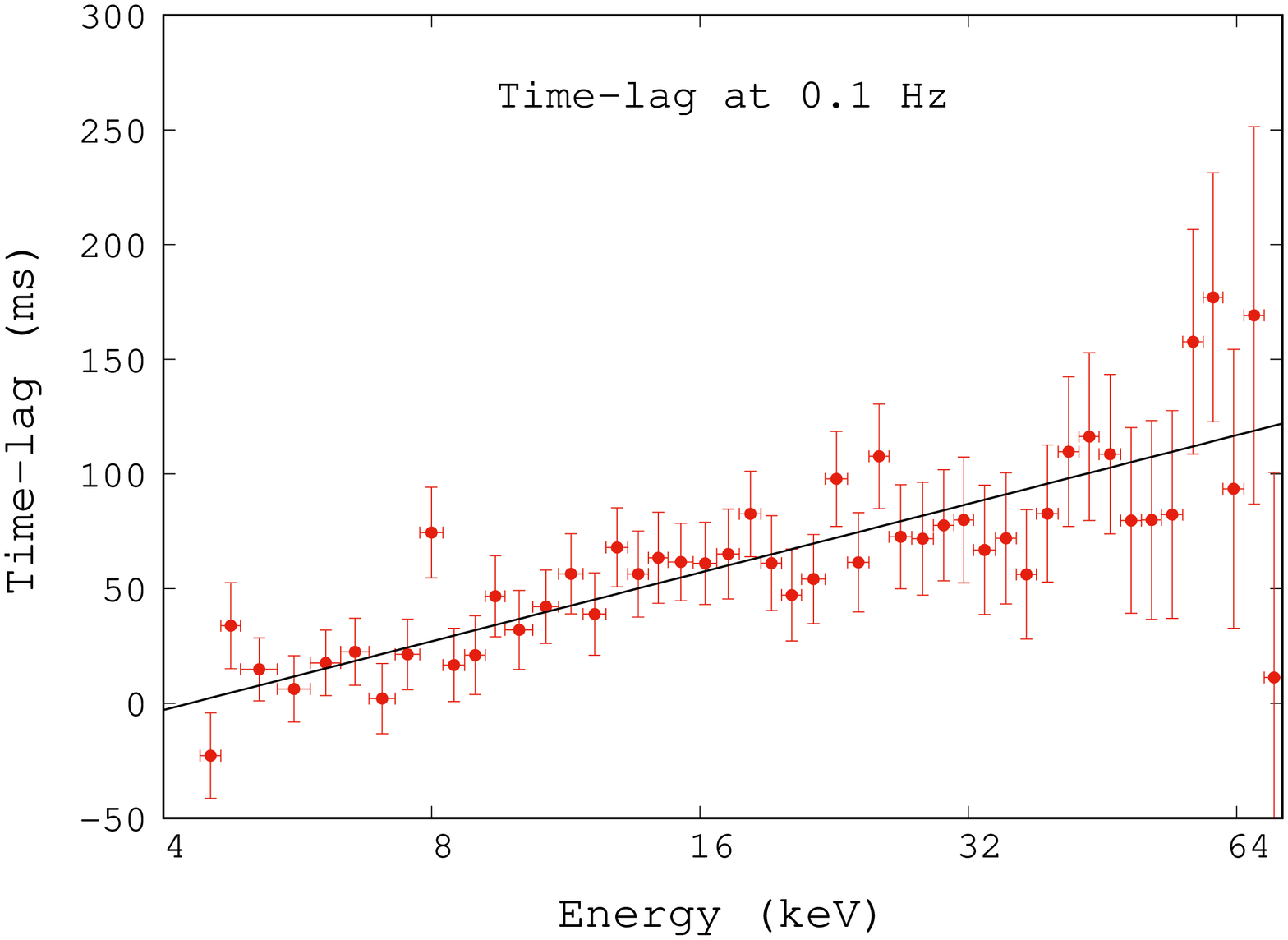}
\includegraphics[width=0.34 \textwidth,angle=-90]{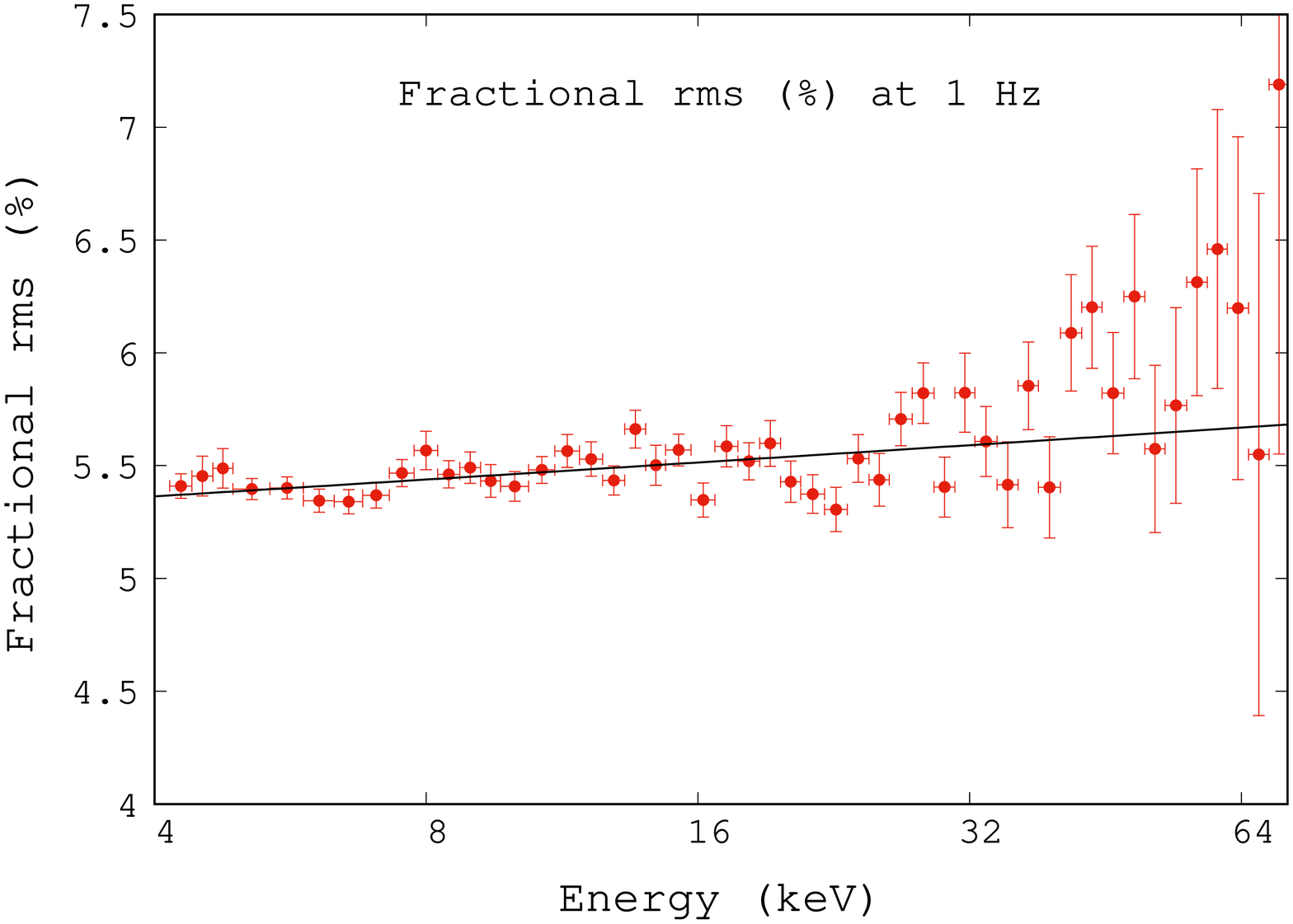}
\includegraphics[width=0.34 \textwidth,angle=-90]{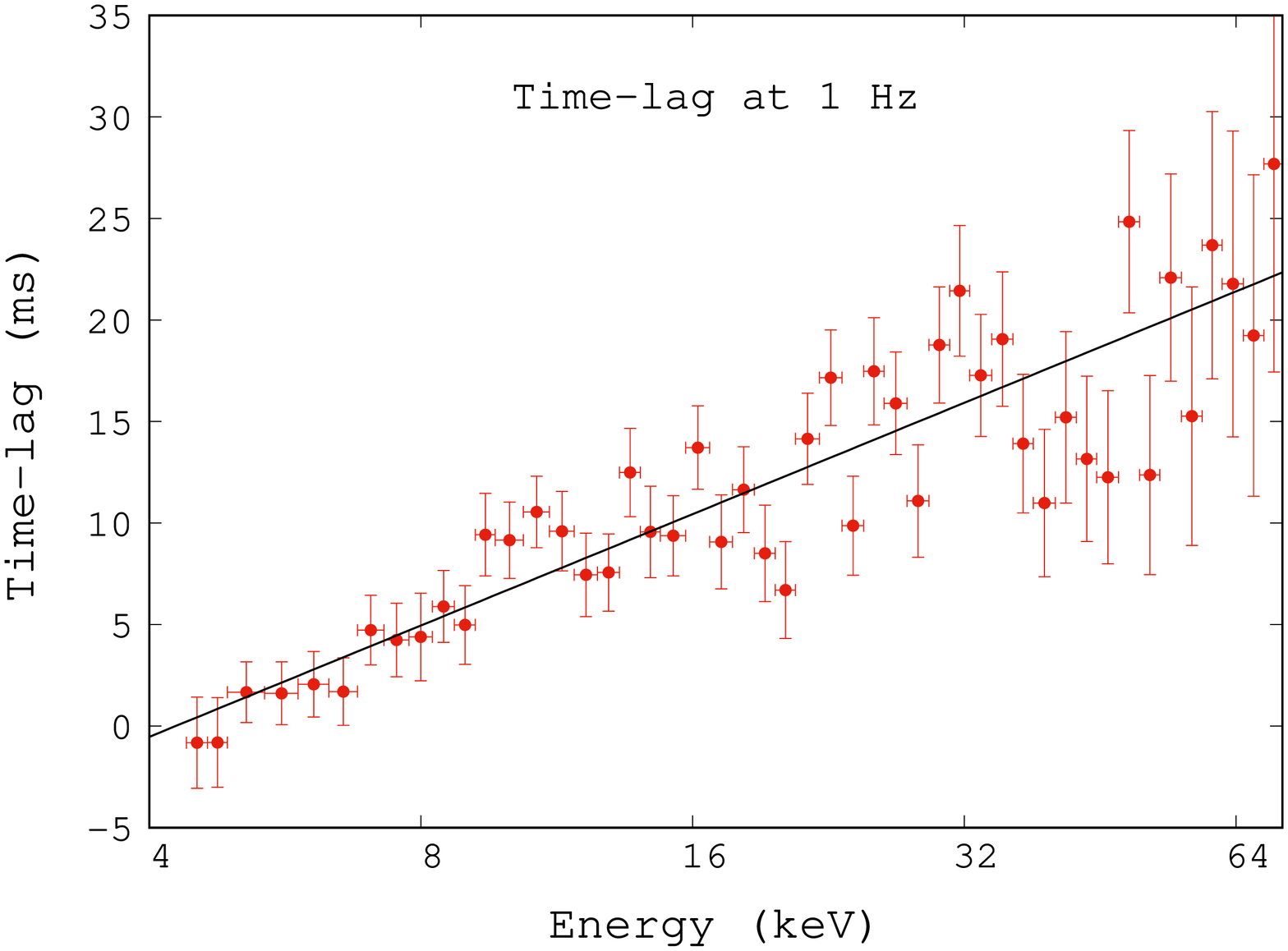}
\includegraphics[width=0.34\textwidth,angle=-90]{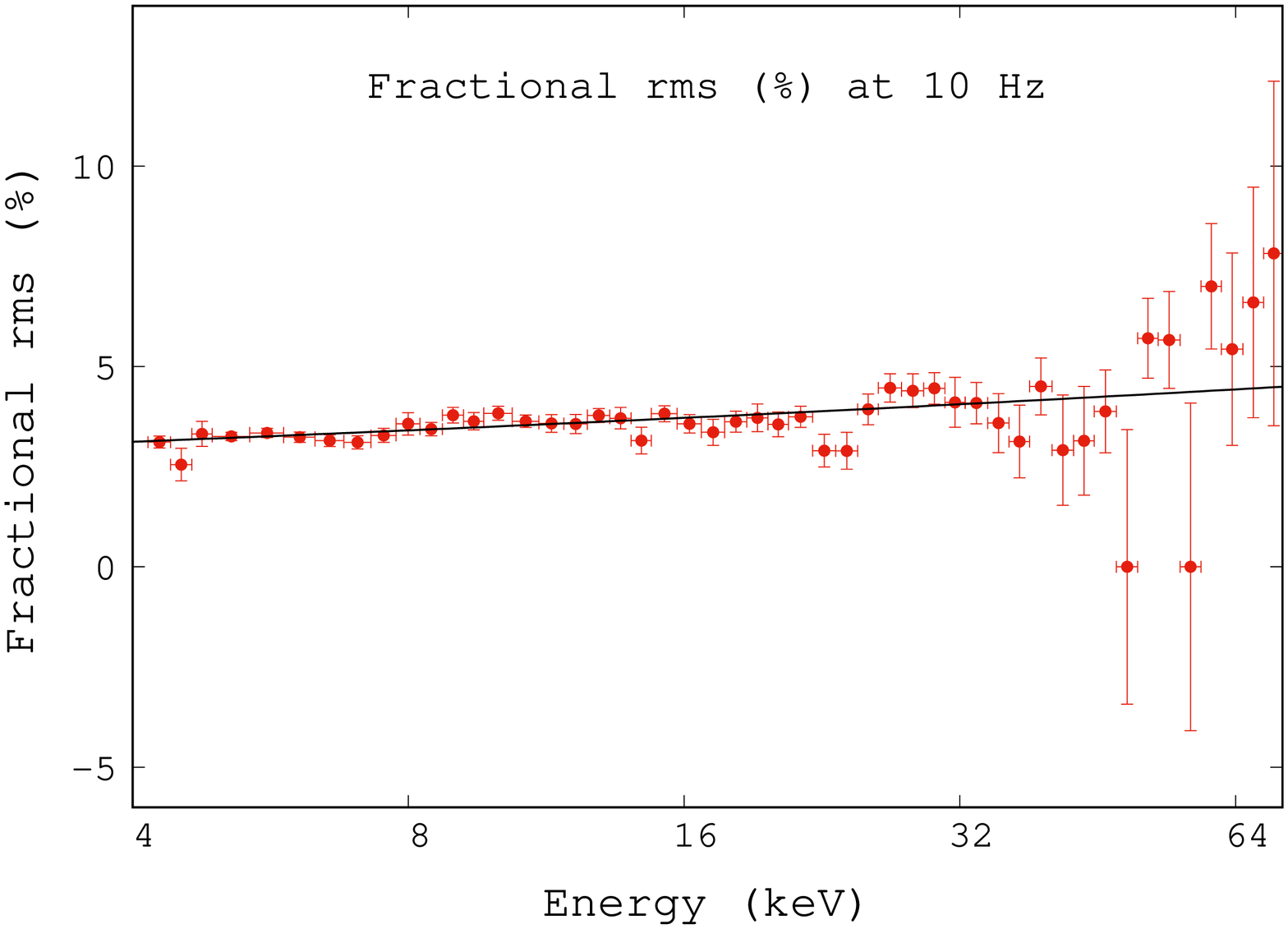}
\includegraphics[width=0.34\textwidth,angle=-90]{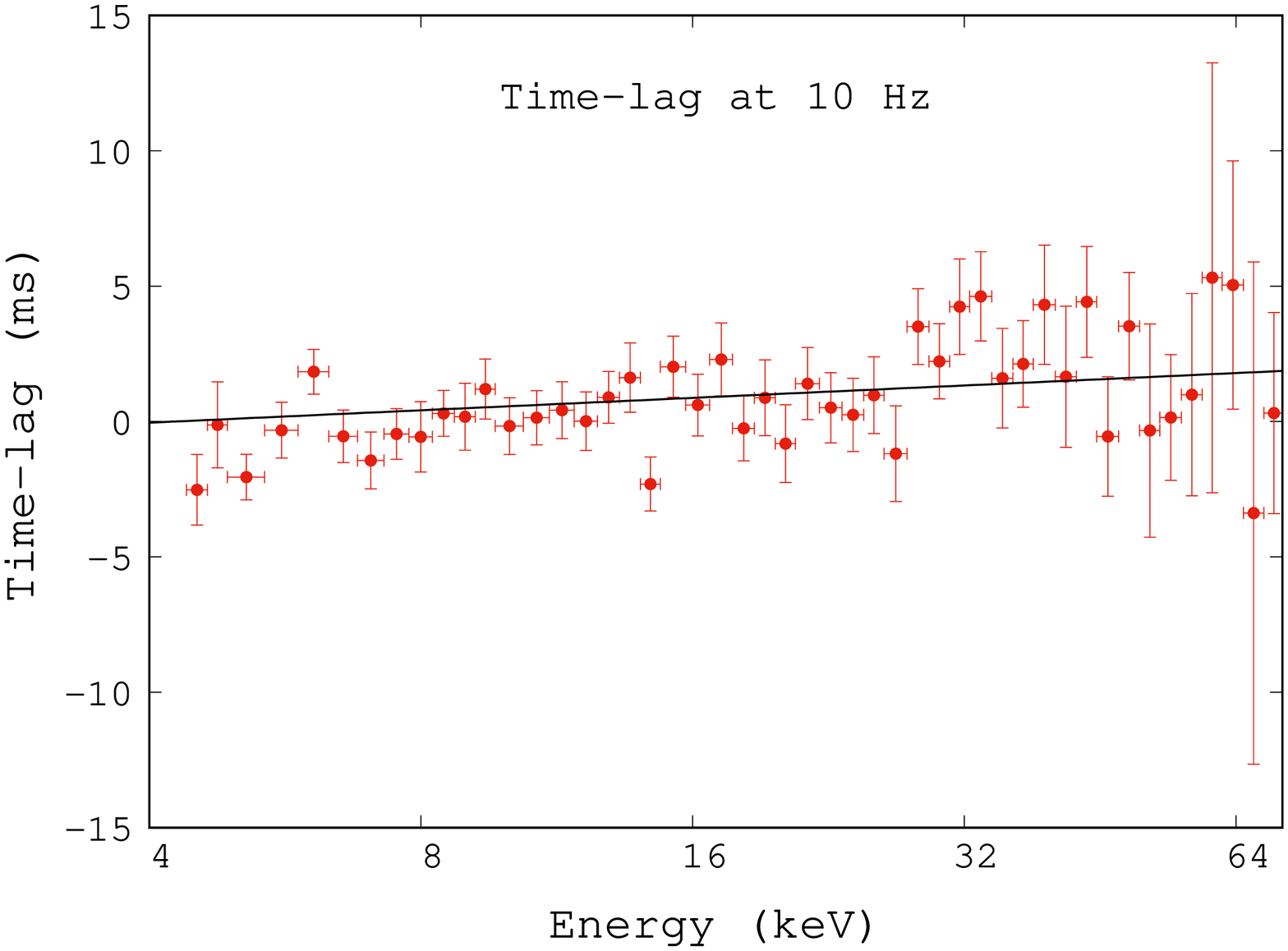}
\caption{Fractional rms and time-lag for three different frequencies 0.1 Hz (top panel), 1 Hz (middle panel) and 10 Hz (bottom panel) for June 2016.}
\label{junrmslag}
\end{figure*}

\begin{figure}
\centering
\includegraphics[width=0.3\textwidth,angle=-90]{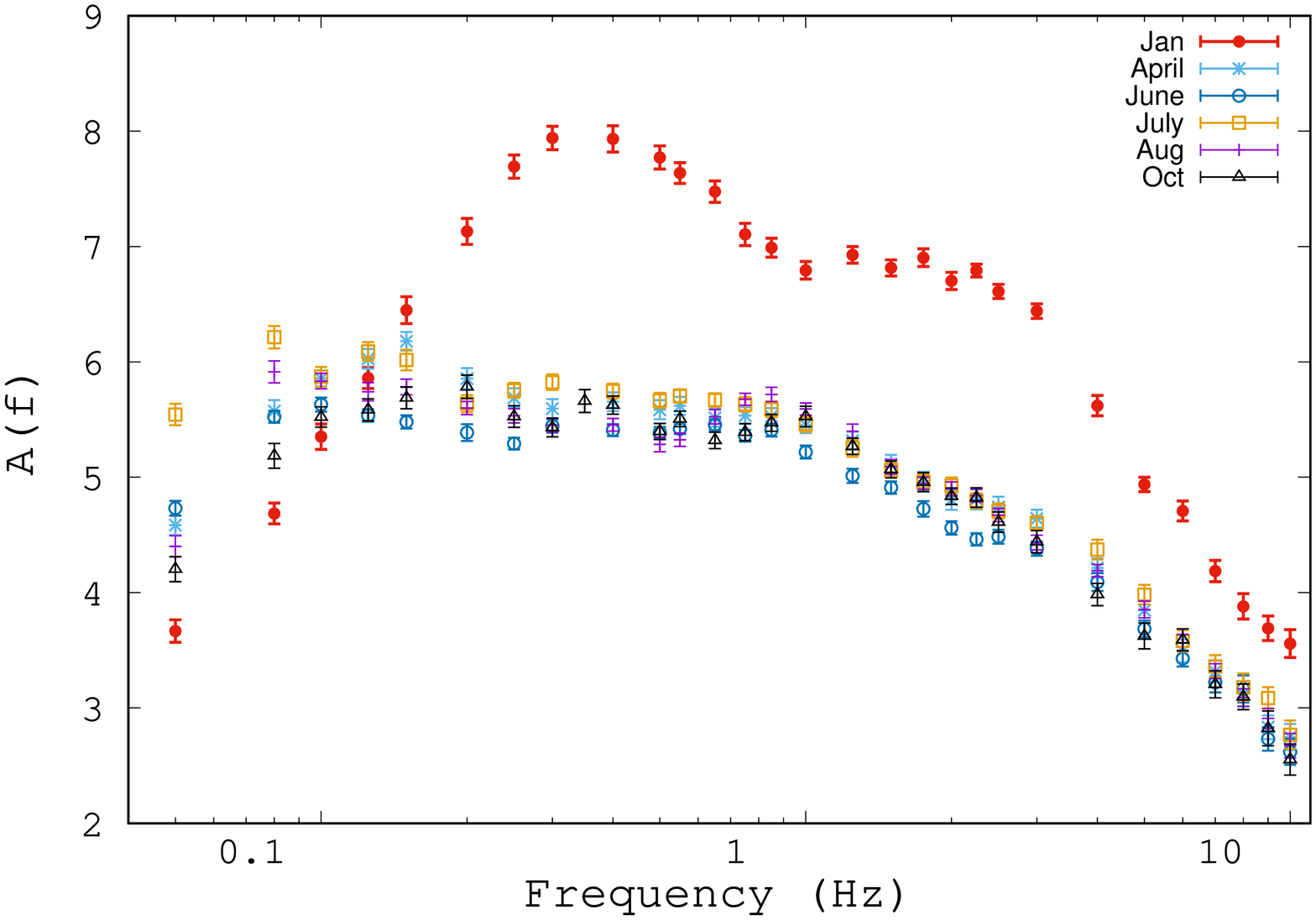}
\includegraphics[width=0.3\textwidth,angle=-90]{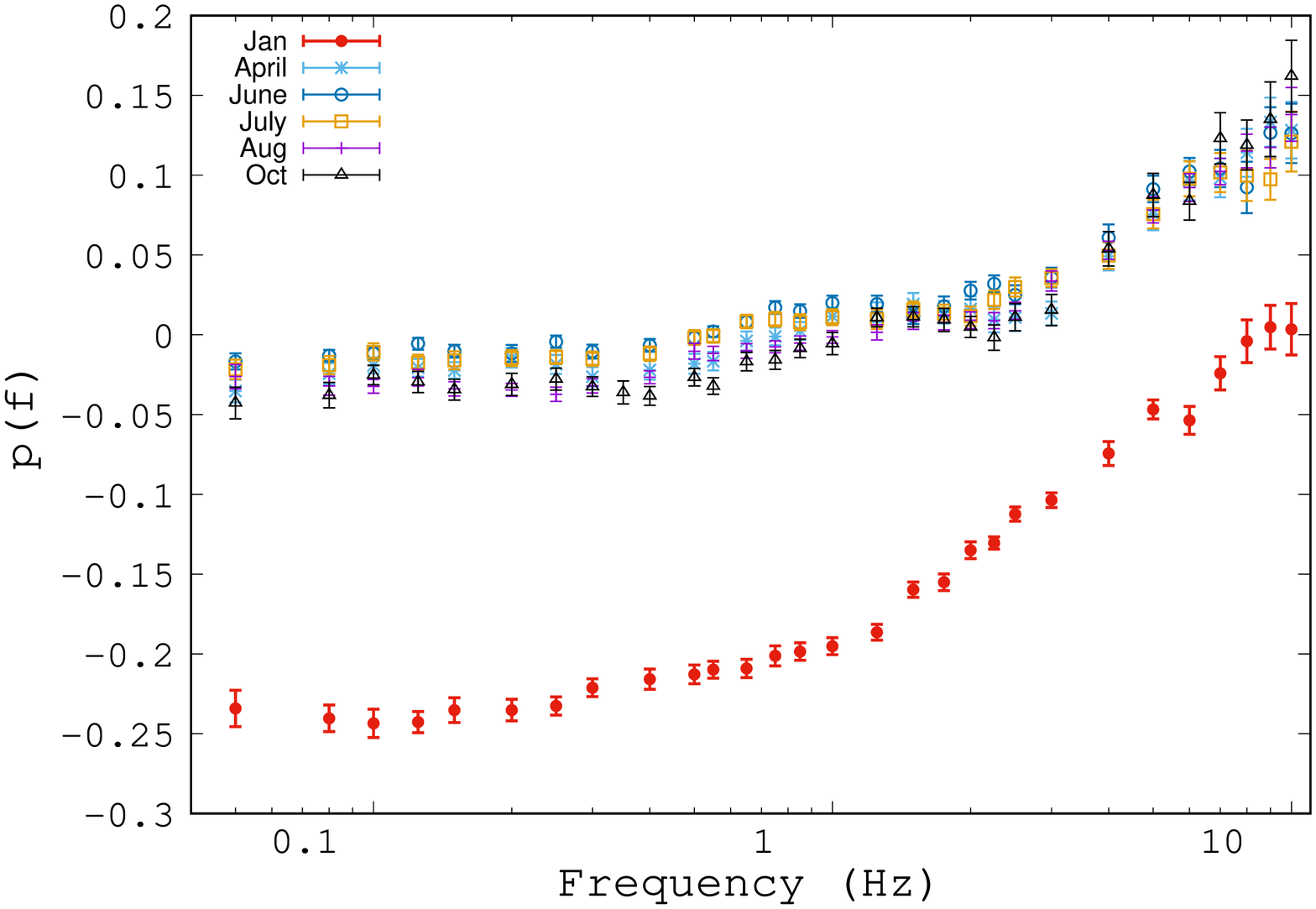}
\includegraphics[width=0.3\textwidth,angle=-90]{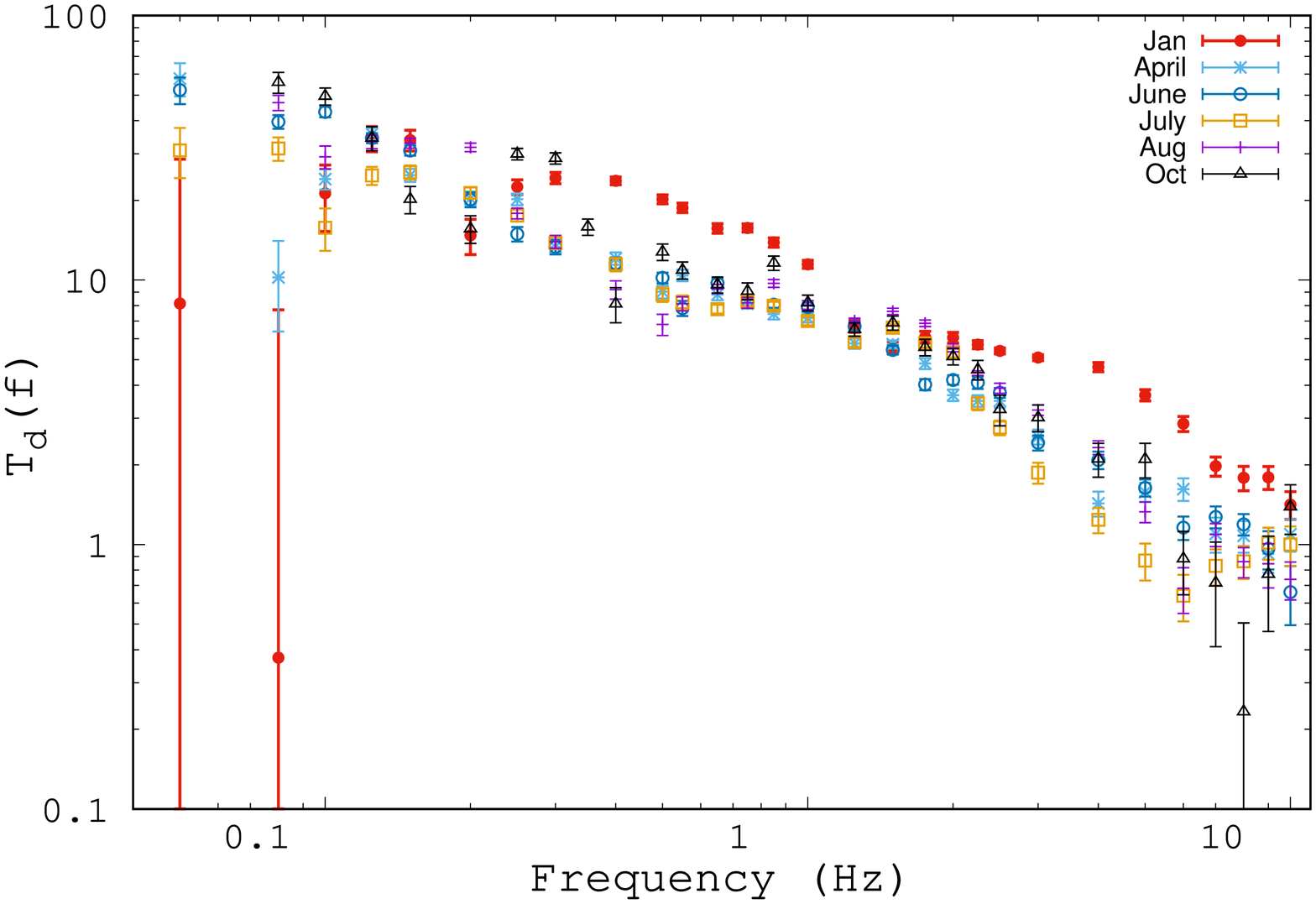}
\caption{Empirical fit parameters \textendash A(f) (Top panel), p(f) (Middle panel) and $T_{d}$(f) (Bottom panel) as a function of frequency} for all the observations.
\label{A_p_comb}
\end{figure}

The standard LAXPC data analysis procedure was used to obtain the PDS in the energy range 4 -80 keV. These spectra have been fitted using representative Lorentzian functions, renormalized and a systematic error of 2 $\%$ added. The PDS of Cygnus X-1 in its hard state is reported to exhibit 2\textendash4 peaks depending upon the frequency range studied and the spectral state of the source \citep{Now00, Pot03, Axe05}. The PDS for January and June data are shown in Figure \ref{powspec}. As reported by \citet{Mis17}, the January data shows only two peaks in the PDS, whereas three peaks are observed for all other observations. Our results are consistent with the results of \citet{Axe08}. For the frequency range up to 10 Hz, only two components (peaks) are present in the PDS and the third component is observed only when the spectrum is relatively harder.
\\[6pt]
Energy dependence of the temporal behavior can be studied by computing the PDS in 3 or 4 broad energy bins as has been reported for {\it RXTE} data \citep{Wil04}. The complete event mode data of LAXPC allows for the fractional rms to be integrated over a certain frequency range with finer energy bins. Moreover, time-lag as a function of these energy bins can also be obtained for a particular frequency. Such an approach has the advantage that the results can be more directly compared with radiative components. The trade off here is that the energy dependent rms and time-lag have to be computed at a frequency, f (Hz) by integrating over a large frequency range to ensure sufficient statistics. Experimenting with different values of $\Delta$f, we found
that taking $\Delta$f $\approx$ 0.4f is optimal as lower values do not constrain the time-lag well and  larger ones do not allow for a meaningful frequency dependence.
\\[6pt]
We have computed time-lags and fractional rms from the time-averaged cross-spectrum using the methods discussed in \citet{Now99} for all the data sets and plotted them as a function of energy for different frequencies. Figures \ref{janrmslag} and \ref{junrmslag} show the energy dependent rms and time-lag for two representative data sets (January and June) at three representative frequencies (0.1, 1 and 10 Hz). The rms either increases or decreases with energy depending on the frequency under consideration, while the time-lag increases with energy indicating that they are hard time-lags where the high energy photons lag the soft ones. The time-lags were computed with respect to the reference energy band 4.15\textendash 4.39 keV.
\\[6pt]
Simple empirical relations can represent the energy dependence of the fractional rms and time-lag. The energy dependency of fractional rms may be represented by 
\begin{equation}\label{eq1}
  {\rm F}(E,f) = A(f)*E^{p(f)}
\end{equation}
where $E$ is the energy (keV), $f$ is the frequency (Hz) under consideration, $A(f)$ and $p(f)$ are constants. Similarly, the energy dependent time-lag may be represented by
\begin{equation}\label{eq2}
\delta t (E,f) = T_d (f)*log (\frac{E}{E_{ref}})  
\end{equation}
where $E_{ref} = 4.27$ keV is the reference energy and $T_d(f)$ is a constant.
We have fitted fractional rms and time-lags versus frequency with the empirical relations in equations (\ref{eq1}) and (\ref{eq2}) respectively to obtain the best fit parameters, $A(f)$, $p(f)$ and $T_{d}(f)$ (Figures {\ref{janrmslag} and \ref{junrmslag})}. The reduced $\chi^2$ was found to be less than 2. Figure \ref{A_p_comb} shows the variation of the best fit parameters, $A(f)$, $p(f)$ and $T_{d}(f)$ as a function of frequency. It is clear from the Figure that while all the other observations show similar behaviour, the January data is distinct. Nevertheless, the time-lag behaviour is similar for all the data sets. 
\\[6pt]
Note that while Figure \ref{A_p_comb}  gives a complete description of the
temporal behaviour of the source in frequency and energy space, it is however
difficult to interpret physically the behaviour of the system and its connection to the time-averaged energy spectra. To make such a connection, a physical model has to be developed and the energy dependent rms and time-lag have to be compared with its prediction. In the next section, we present one such model.

\subsection{Energy dependent time-lag and fractional rms from a inner hot disk}
\begin{figure*}
\centering
\includegraphics[width=0.34\textwidth,angle=-90]{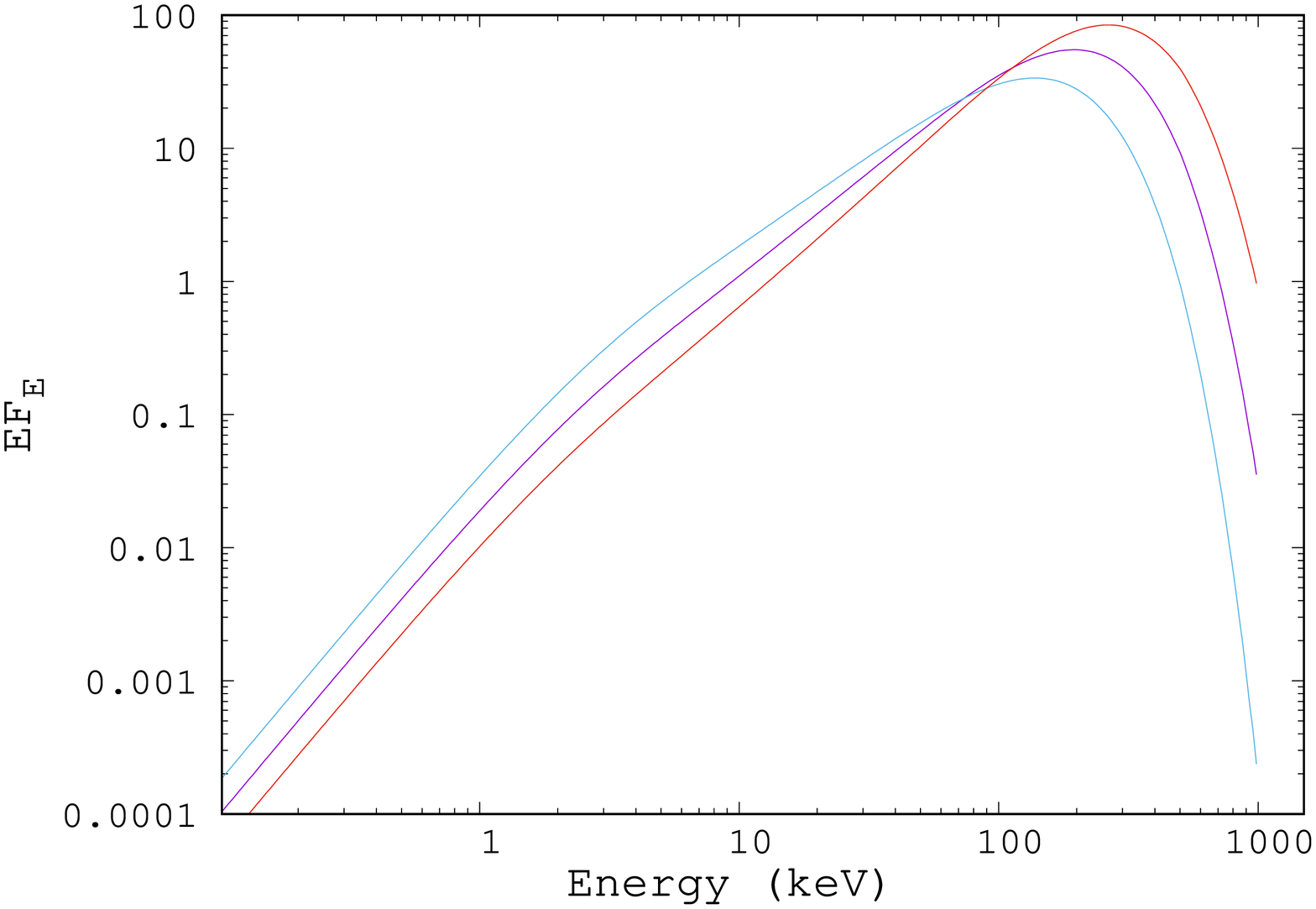}
\includegraphics[width=0.34\textwidth,angle=-90]{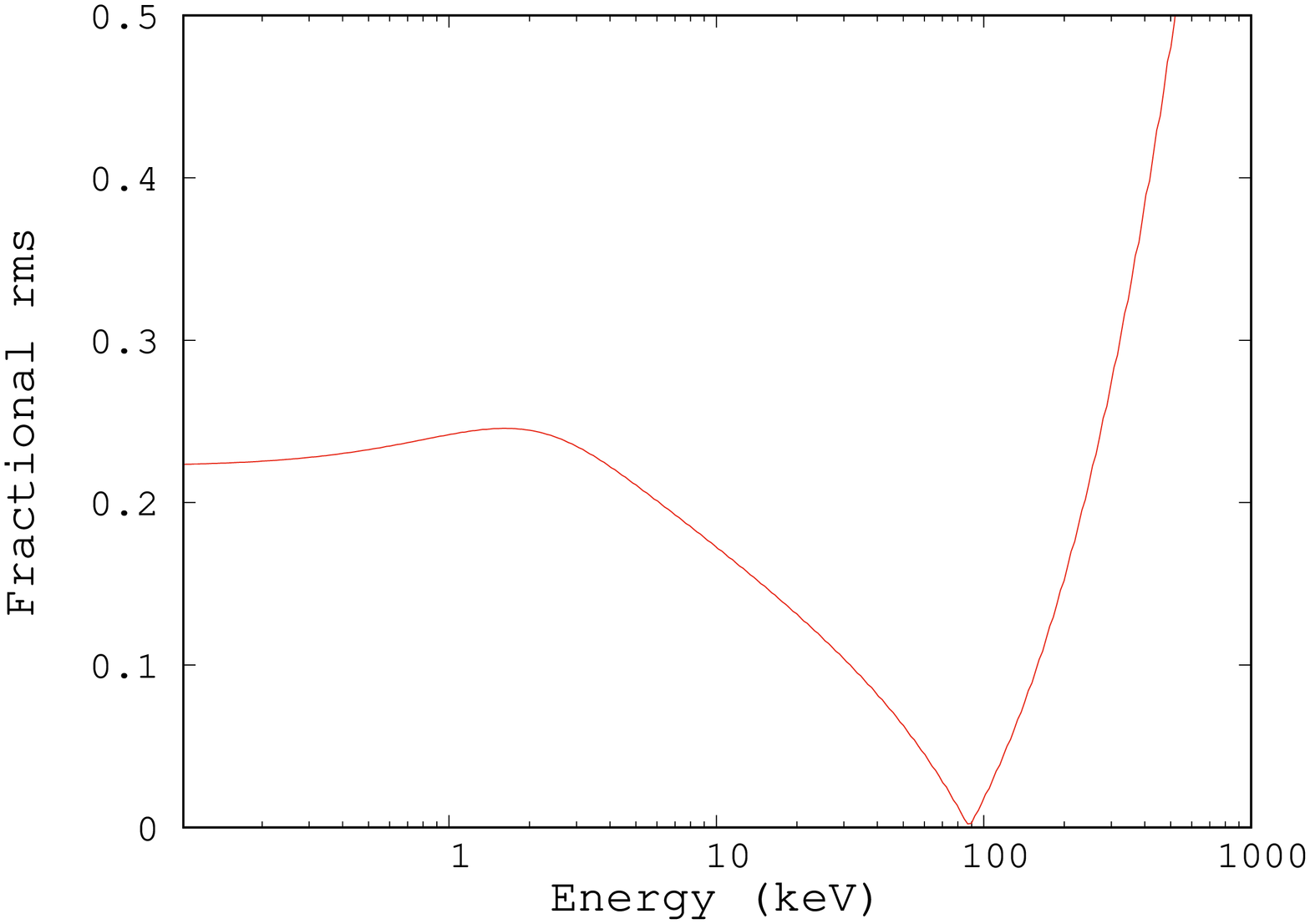}
\includegraphics[width=0.34\textwidth,angle=-90]{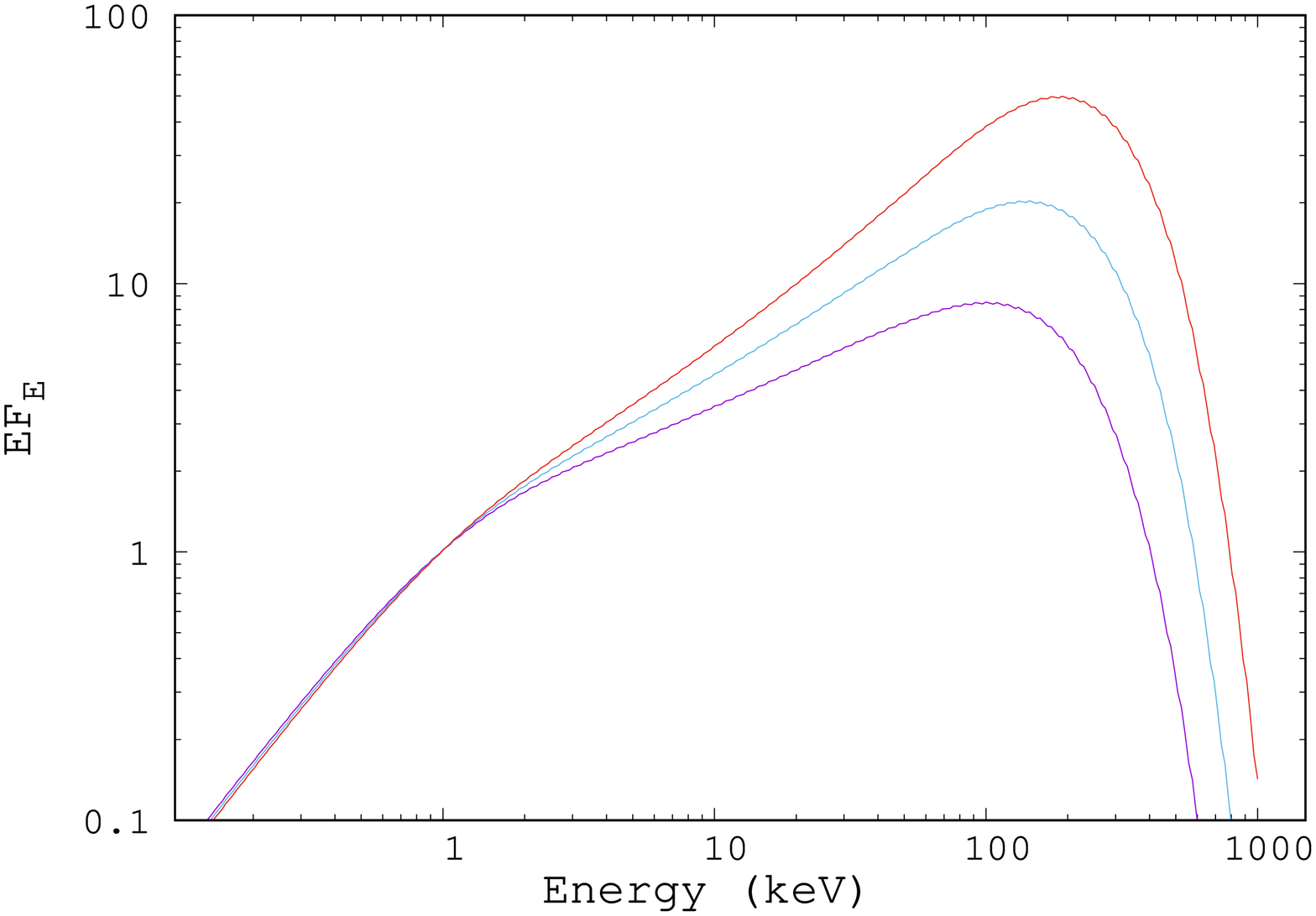}
\includegraphics[width=0.34\textwidth,angle=-90]{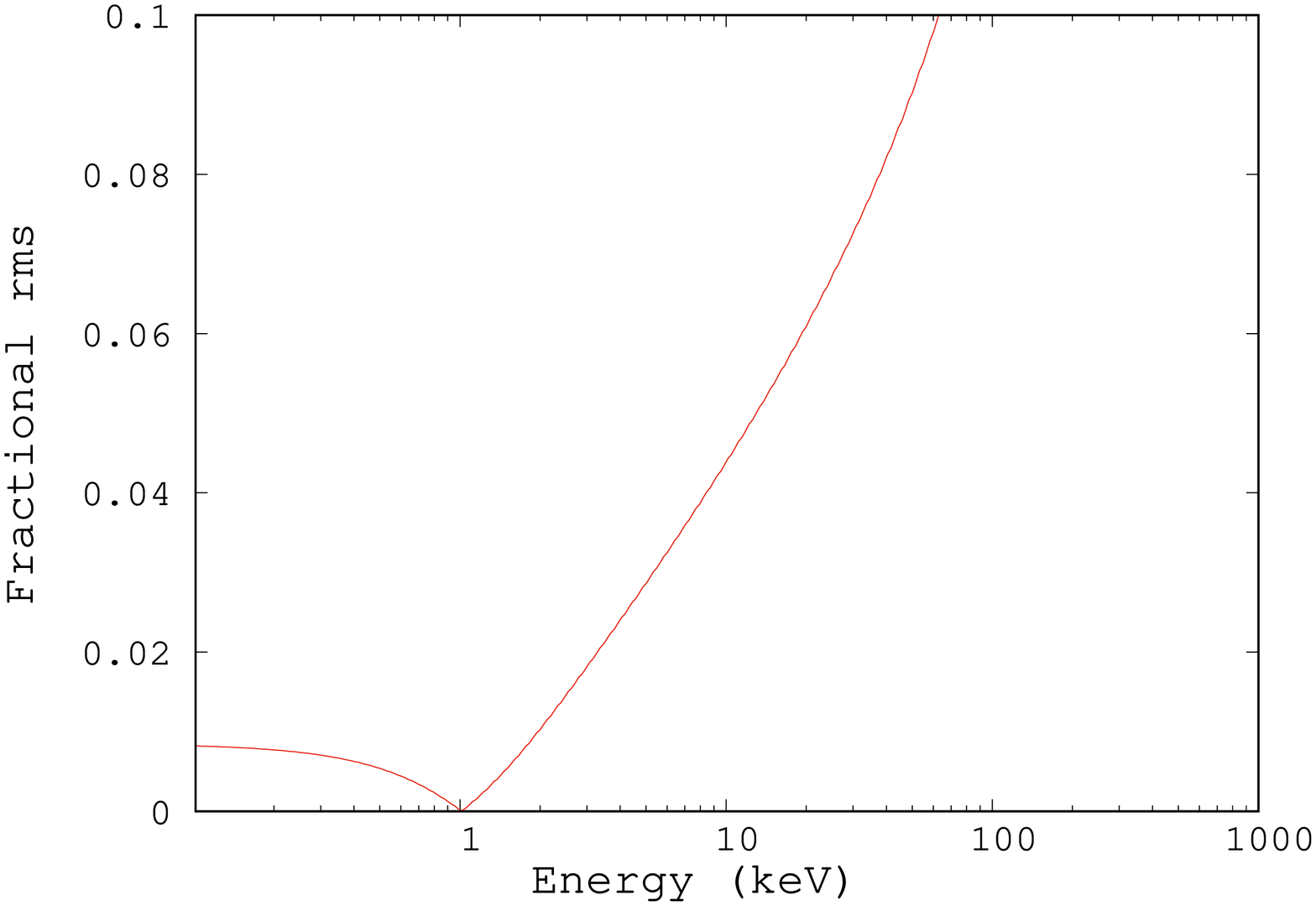}
\caption{Figures to illustrate the response of the Comptonized spectrum to variation in the seed photon temperature, T$_{s}$ (Top left  panel) and the electron temperature, T$_{e}$ (Bottom left panel). The top left panel shows three spectra with different seed photon temperature showing that the spectra pivot at high energy ($\sim 90$ keV) which reflects in the fractional rms (Top right panel) decreasing with energy in the LAXPC band. On the other hand, for the electron temperature variation, the three representative spectra on the bottom left panel show pivoting at low energy ($\sim 1.0$ keV) causing the fractional rms to increase with energy in the same band (Bottom right panel).}
\label{theoryvar1}
\end{figure*}
\begin{figure*}
\centering
\includegraphics[width=0.34\textwidth,angle=-90]{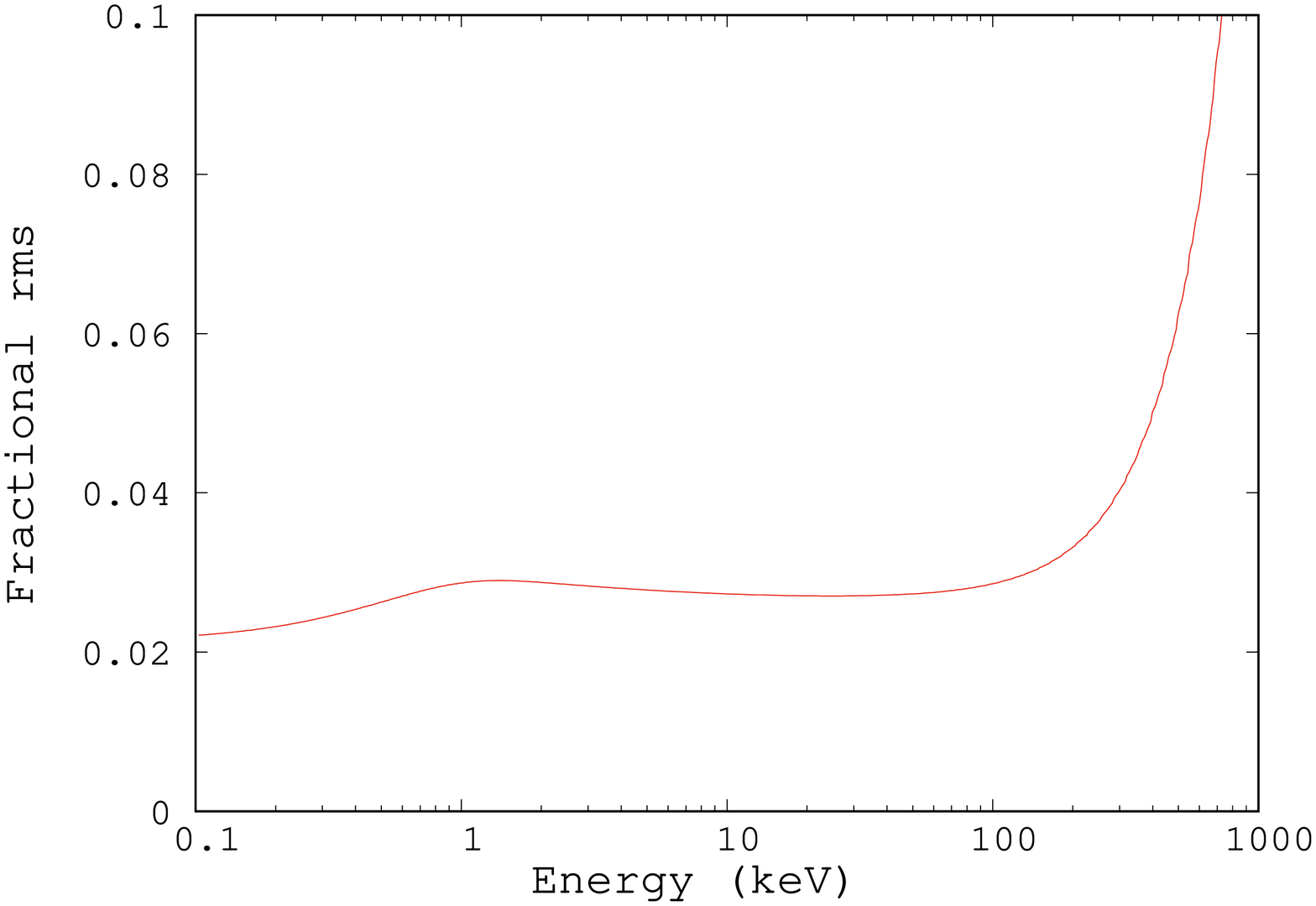}
\includegraphics[width=0.34\textwidth,angle=-90]{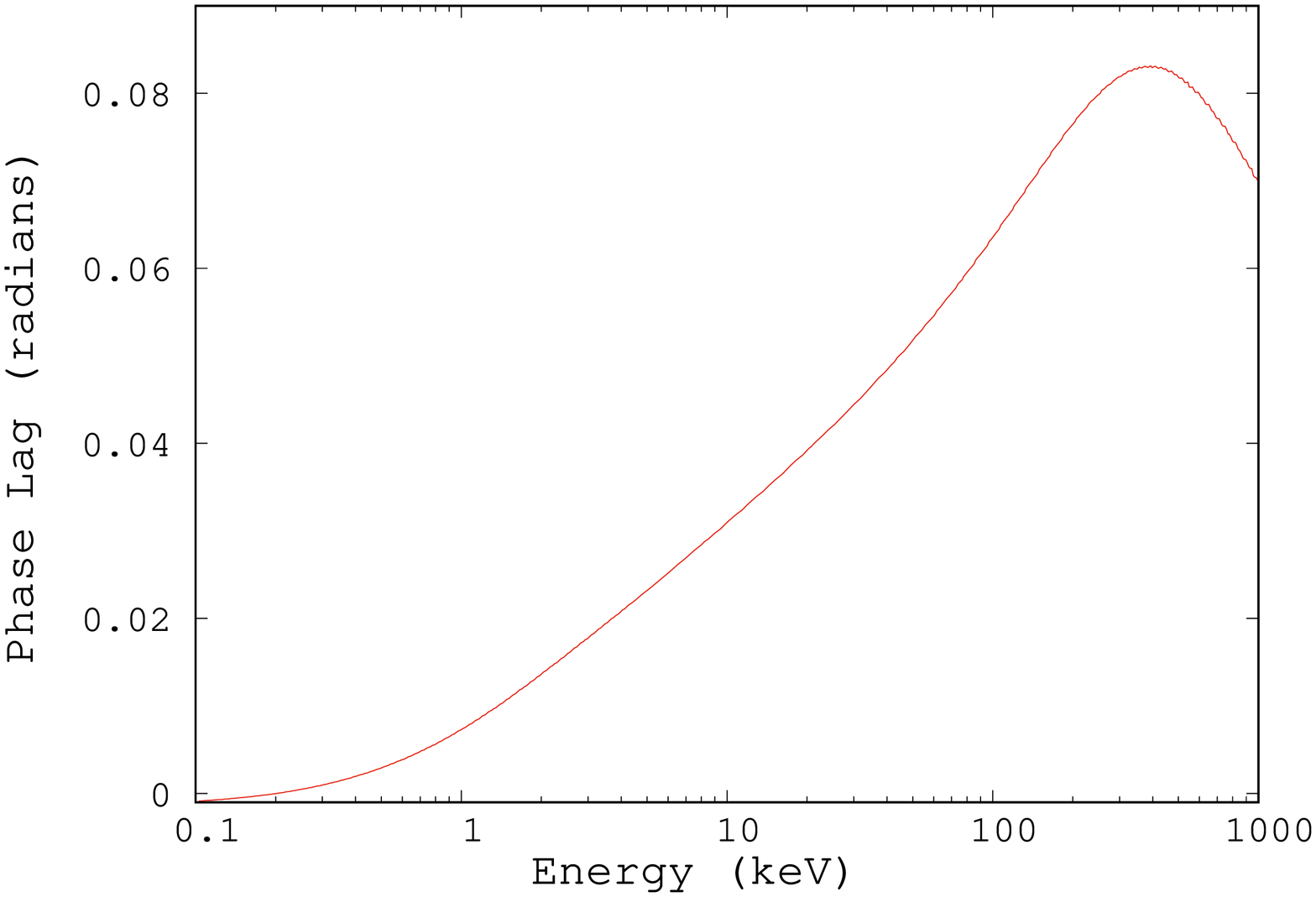}
\caption{Example of the response of the spectrum to coherent variations in the seed photon and inner disk temperatures. The fractional rms can increase/decrease with  energy (Left panel), while
phase-lags can occur between two energy bands for non zero time delay between the temperature variations (Right panel). The parameters used for these plots are fiduciary.}
\label{theoryvar2}
\end{figure*}
We consider the geometry of the hard state of Cygnus X-1 to be that of a standard accretion disk truncated at a large radius with an inner hot accretion disk. The inner disk which is assumed to be homogeneous and having a uniform temperature $T_{e}$, Comptonizes the seed photons arising from the outer cold standard disk producing the primary emission observed by the LAXPC in the energy range 4\textendash 80 keV. Thus, the Comptonized spectrum, $S_{c}(E, T_{e}, T_{s})$ is characterized by the input soft photon source $S_{s}(E, T_{s})$, the optical depth $\tau$ and electron temperature $T_{e}$, where $T_{s}$ is the seed photon temperature. The spectral shape of the soft photon source is that of a truncated standard disk and its luminosity $L_{s}$ is proportional to $R_{T}^2T_{D}^4$, where $R_{T}$ is the truncation radius and $T_{D}$ is the temperature at that radius. \\[6pt]
In the stochastic propagation model, fluctuations occur at outer regions of the disk which then propagate inwards. This will lead to variations of the temperature ($\Delta T_s$) of the truncation radius and the flux of photons entering the inner hot region. The change in the input flux will lead to a corresponding change in the temperature of the inner disk to maintain the same power output. Subsequently after a propagation time delay, the perturbation will reach the inner regions causing a variation of its temperature $ \Delta T_{e}$. In the first order, the variation of the Comptonized spectrum due to these effects can be written as
 \begin{multline}
\Delta S_c(E, T_{e}, T_{s}) = \\ \frac{\partial{S_{c}(E, T_{e}, T_{s})}}{\partial{T_{s}}} \Delta T_{s} (t) + \frac{\partial{S_{c}(E, T_{e}, T_{s})}}{\partial{T_{e}}} \Delta T_{e} (t-\tau_{D}) 
\end{multline}
where $\tau_D$ is the time taken for the perturbation to propagate from the truncation radius to the inner region. The partial derivative with respect to seed photon temperature ($\frac{\partial{S_{c}(E, T_{e}, T_{s})}}{\partial{T_{s}}}$) has
to be computed at constant heating rate of the hot region and it is assumed
that the optical depth of the inner region ($\tau$), the
truncation radius ($R_{T}$) and the fraction of photons that enter the inner
region are all invariant. To obtain the behaviour at a particular frequency, one can introduce $\Delta T_{s} (t) = |\Delta T_{s} (t)|e^{i2\pi ft}$ and
 $\Delta T_{e} (t) = |\Delta T_e (t)|e^{i2\pi f(t-\tau_D)}$. The corresponding fractional rms with energy is then given by $(1/\sqrt{2})|\Delta S_{c}(E, T_{e}, T_{s})|/S_{c}(E, T_{e}, T_{s})$ and
the phase-lag between the variability at $E$ and $E_{ref}$ will be the phase of $\Delta S_{c}^*(E_{ref}, T_{e}, T_{s})\Delta S_{c}(E, T_{e}, T_{s})$.
\\[6pt]
We numerically obtain the partial derivatives using the modified Kompaneets equation \citep{Kom56} as formulated in the Fortran code used by the  XSPEC model `{\it nthcomp}'. We first compute the time averaged Comptonized spectrum, $S_{ca}$(E, $T_{ea}, T_{sa})$ based on the best fit spectral parameters (the temperature, $kT_{ea}$, the optical depth 
$\tau_{a}$ and the seed photon temperature of the disk emission $kT_{sa}$) as obtained in \S 2.1\footnote{The Xspec routine '{\it nthcomp}' has the parameters as photon index $\Gamma$ instead of the optical depth $\tau$ which is convenient for spectral fitting. The relation between $\Gamma$, $\tau$ and the temperature is as follows. \begin{equation}
\Gamma = [(9/4)+ (3m_{e}c^{2}/{kT_{e}*((\tau+3/2)^{2}-9/4}))]^{1/2} - 1/2 \nonumber
\end{equation} \citep{Zdz96, Zyc99, Wilk15}}. The heating rate of the hot disk is computed by subtracting the power of the input soft photon source from the total power i.e., 
\begin{equation}
\dot H = \int E (S_{ca} (E,T_{ea}, T_{sa}) - S_{sa} (E,T_{sa})) dE 
\end{equation}
where $S_{sa} (E,T_{sa})$ is the time averaged input soft seed photon spectrum.  We then consider a variation of the seed photon temperature $T_{s} = T_{sa}+\Delta T_{s}$ and obtain the emergent Comptonized spectrum $S_{c}(E, T_{e}, T_{s})$. The temperature of the inner disk is then varied to ensure that its heating rate remains the same, i.e. $kT_{e}$ is found such that the following equation is satisfied\footnote{In the Xspec routine {'\it nthcomp}', there is an inbuilt re-normalization that sets the output spectrum to 1 at 1 keV. This has been disabled for this exercise.},
 \begin{equation}
\dot H = \int E(S_{c} (E, T_e, T_{s}) - S_{sa} (E,T_{sa})) dE 
\end{equation}
The required derivative can now be computed numerically to be
\begin{equation}
\frac{\partial{S_{c}(E, T_{e}, T_{s})}}{\partial{T_{s}}} \sim \frac{S_{c}(E,T_{e},T_{s}) - S_{ca}(E,T_{ea},T_{sa})}{\Delta T_{s}}  
\end{equation}
for small $\Delta T_{s}$ i.e. $\Delta T_{s} << T_{s}$.
The partial derivative with respect to the inner disk temperature($\frac{\partial{S_{c}(E, T_{e}, T_{s})}}{\partial{T_{e}}}$) can be obtained by subtracting the time-averaged spectrum from the spectrum computed at a slightly higher temperature
$T_{e} = T_{ea}+\Delta T_{e}$ and dividing it by $\Delta T_{e}$. Here all the other parameters are kept constant. 
\\[6pt]
Figure \ref{theoryvar1} shows the response of the Comptonized spectrum when either the seed photon temperature or the inner disk temperature (electron temperature) is varied. The responses are qualitatively different for the two. For the seed photon temperature variation, the spectrum pivots at high energies and hence the fractional rms decreases with energy in the LAXPC band (Top panel of Figure \ref{theoryvar1}). On the other hand, variation of the inner disk temperature causes the spectrum to pivot at low energies giving rise to a fractional rms which increases with energy (Bottom panel of Figure \ref{theoryvar1}). When only one of the temperatures is varied, the response is frequency independent and there is no energy dependent phase-lag, since in this model, the time delays due to light travel time effects  have been neglected.
\\[6pt]
When there is coherent variation of the soft photon and inner disk temperatures, the energy dependent variability of the Comptonized spectrum is more versatile. The fractional rms can either decrease
or increase with energy depending on the relative strengths of the temperature variation and the time delay between the two gives rise to energy dependent time-lag. For fiduciary values, Figure  \ref{theoryvar2} shows the fractional rms and the phase-lag as a function of energy. In the next section, we formally fit the observed energy dependent fractional rms and phase-lag using this model.
\subsection{Fitting the energy dependent temporal properties of Cygnus X-1}
\begin{figure}
\centering
\includegraphics[width=0.34\textwidth,angle=-90]{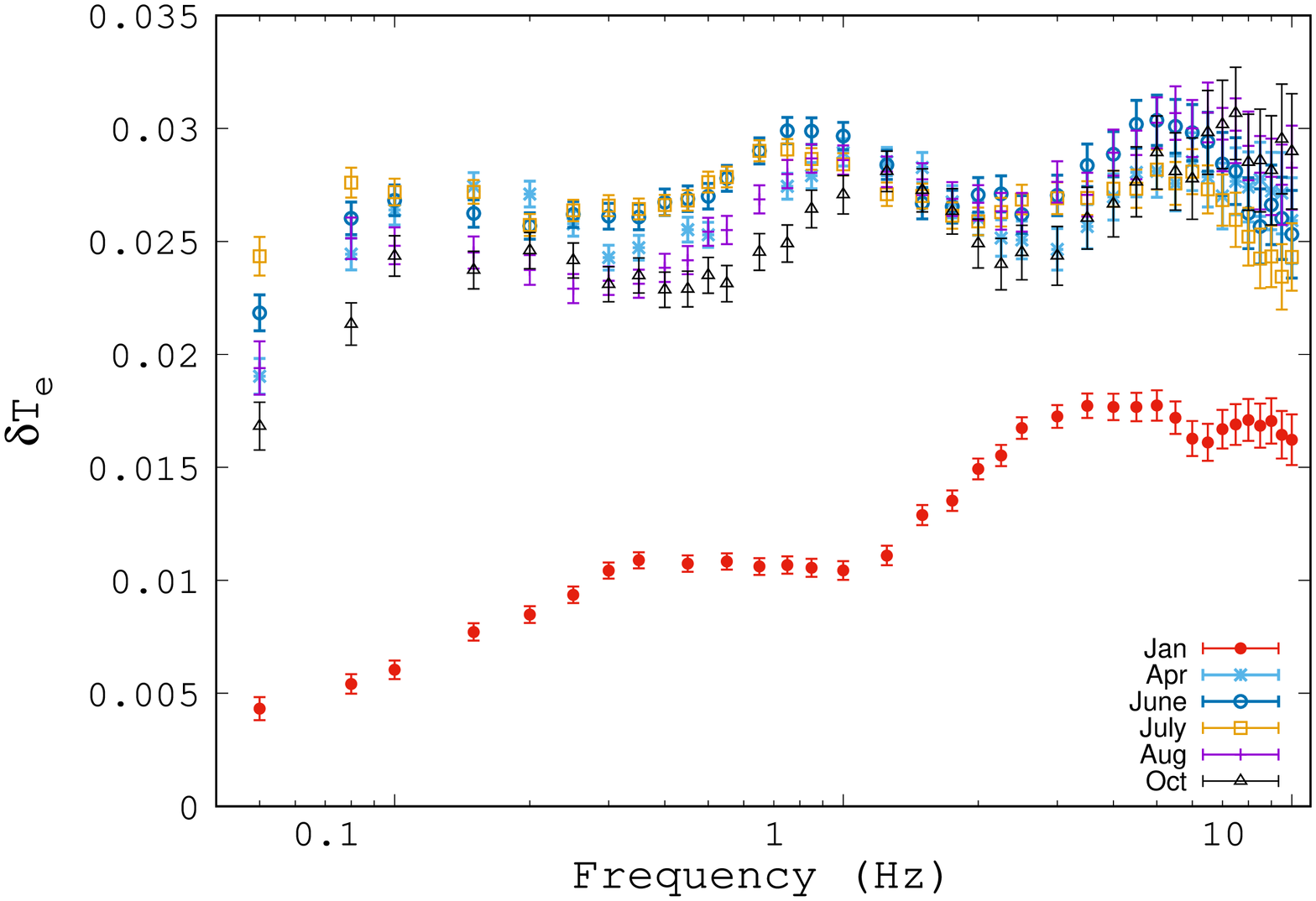}
\includegraphics[width=0.34\textwidth,angle=-90]{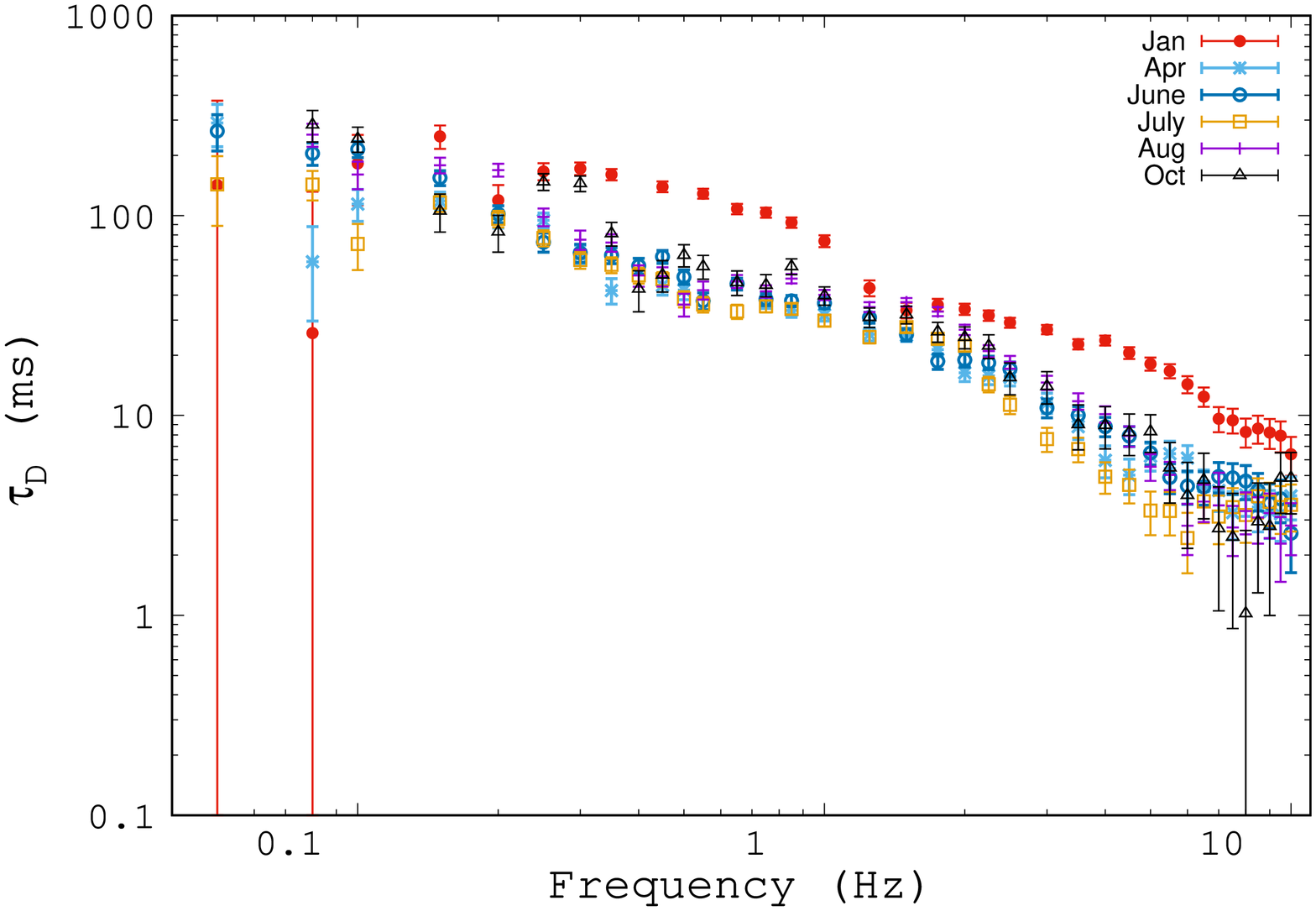}
\includegraphics[width=0.34\textwidth,angle=-90]{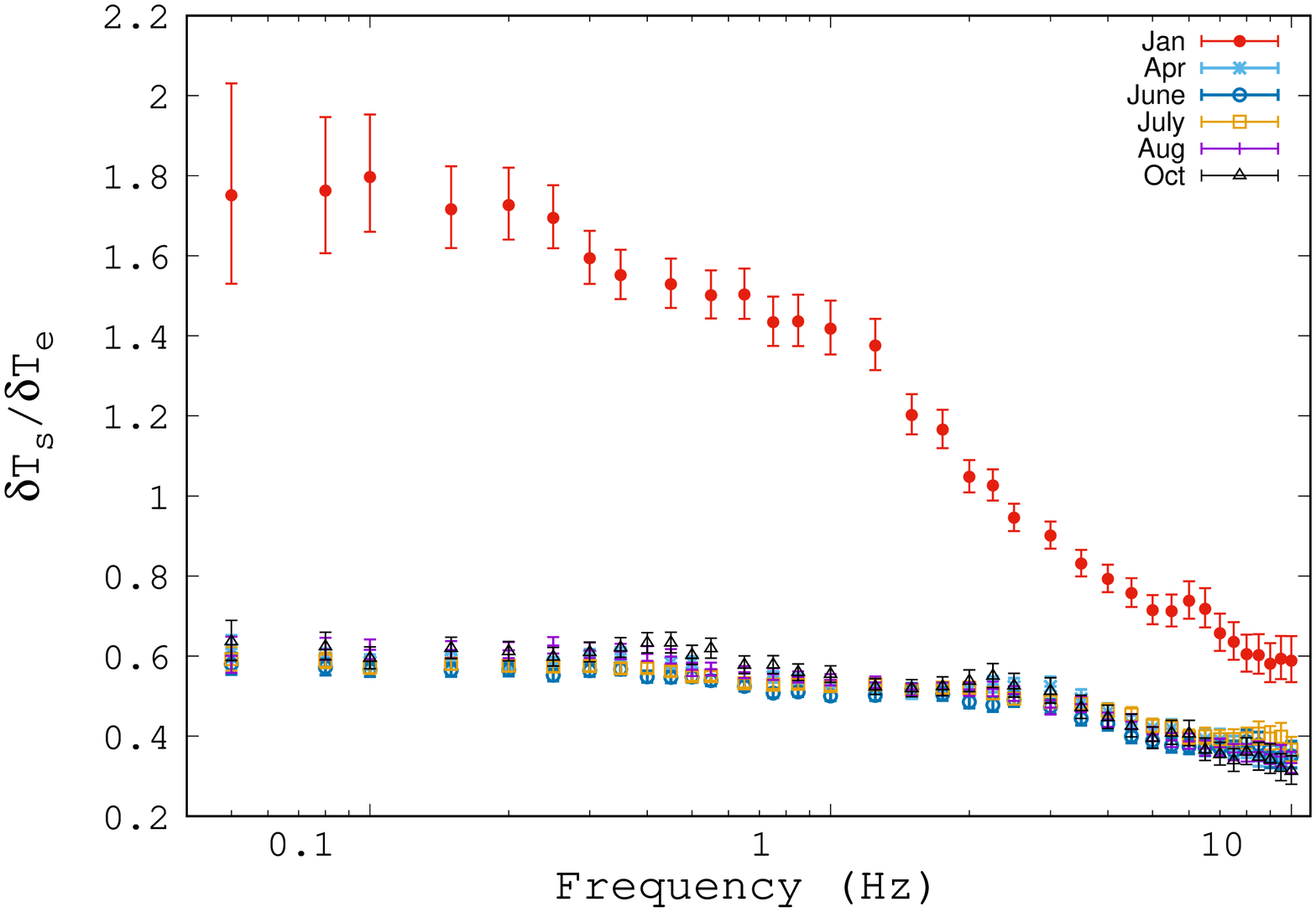}
\caption{Stochastic propagation model fit parameters $\delta T_e$ (Top panel) $\tau_D$ (Middle panel) and $\delta T_s/\delta T_e$ (Bottom panel) plotted as a function of frequency}
\label{model_var_fig}
\end{figure}
\begin{figure}
\centering
\includegraphics[width=0.34\textwidth,angle=-90]{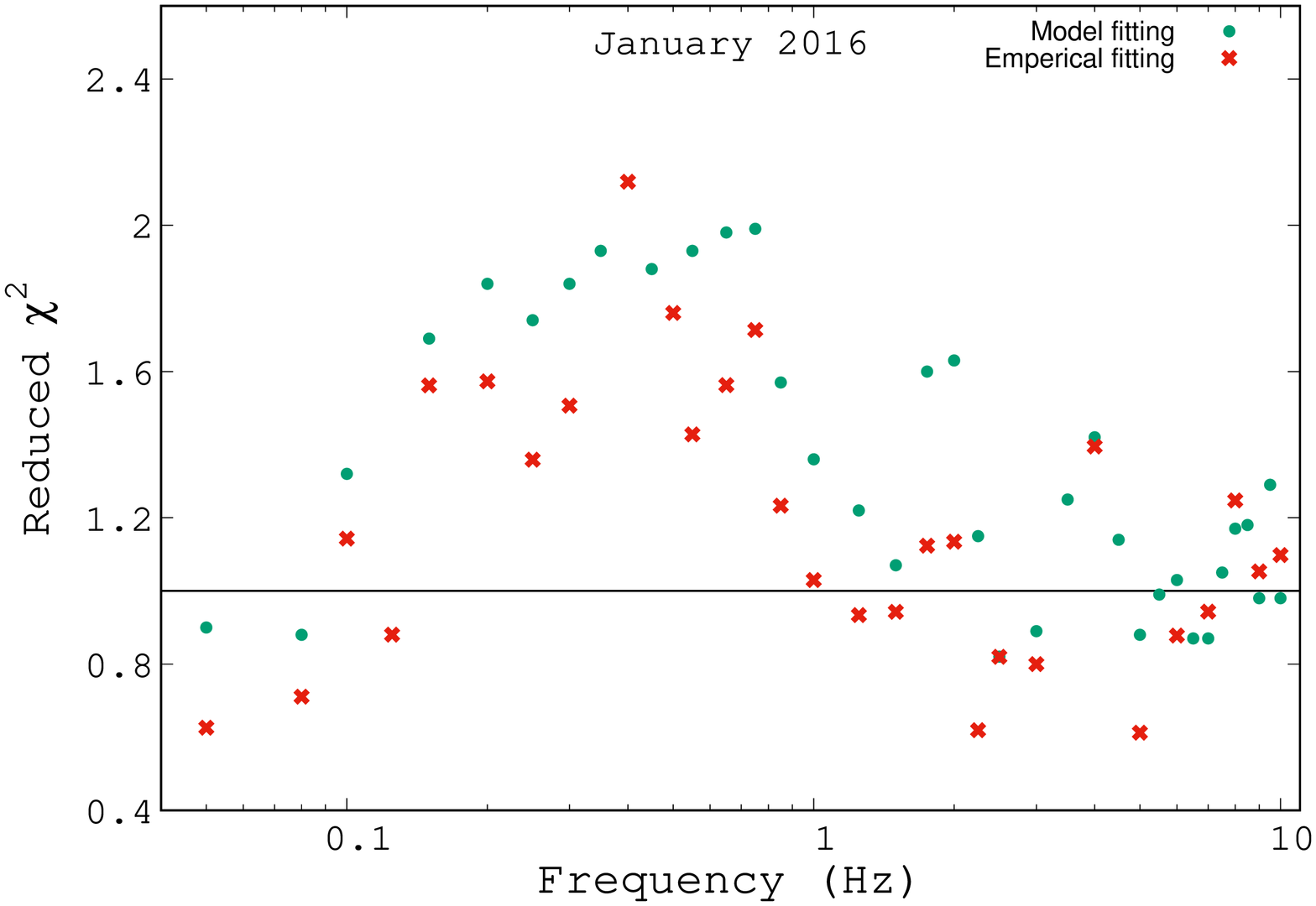}
\includegraphics[width=0.34\textwidth,angle=-90]{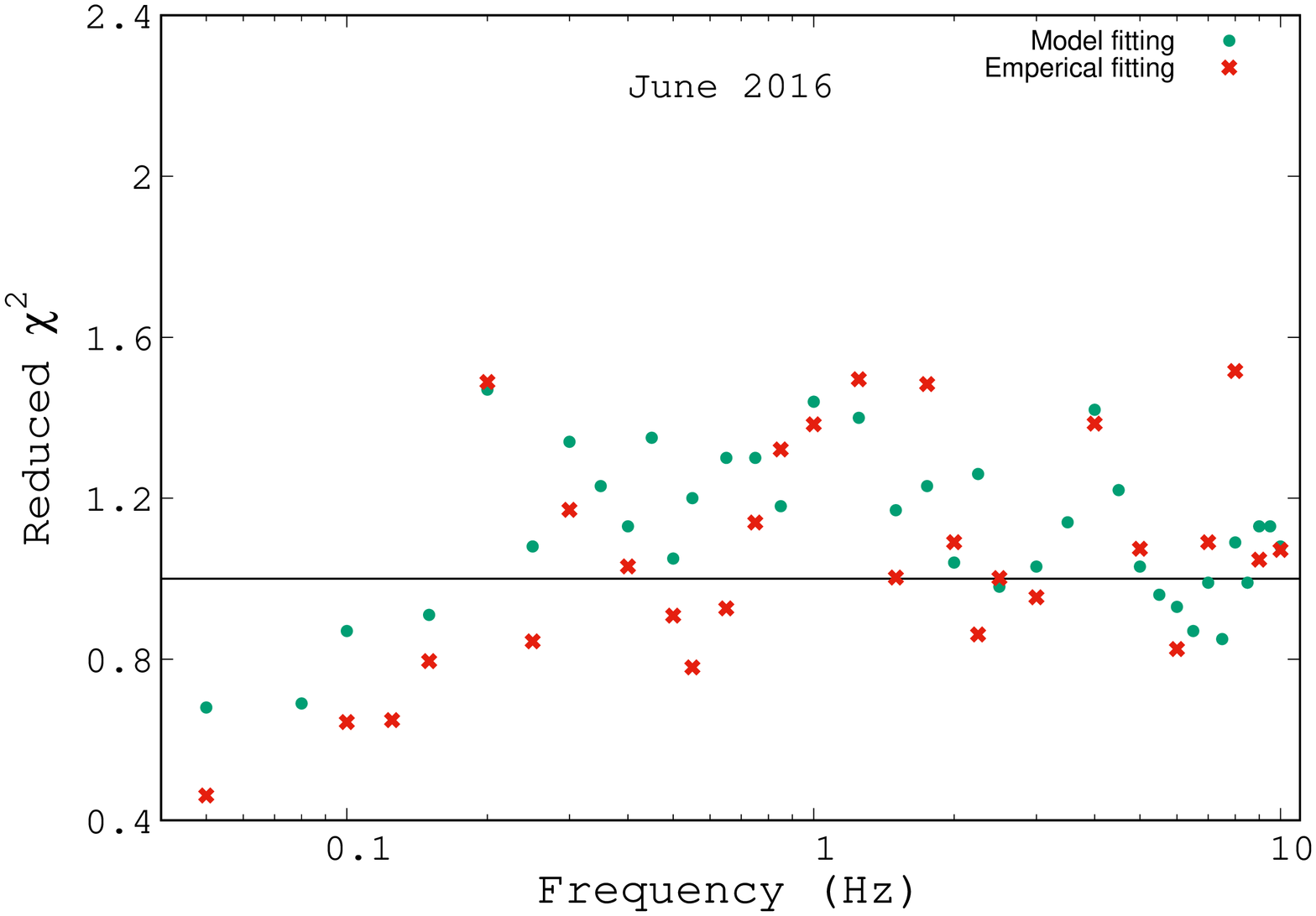}
\caption{Comparison between the model and empirical fits for time-lags as a function of frequency for January (Top panel) and June (Bottom panel)}
\label{redchi1}
\end{figure}

The motivation here is to quantitatively model the energy and frequency dependent fractional rms and phase-lag measured by LAXPC as described in \S 2.2. We associate the variability only to the dominant thermal Comptonization spectrum and neglect the  disk and reflection components in the LAXPC in the energy range 4\textendash 80 keV band. The time-averaged spectrum is characterized by the electron temperature, the optical depth and the soft photon temperature which have been obtained from the spectral fitting described in \S 2.1. 
\\[6pt]
Once the time-averaged spectral parameters are known, the energy dependent fractional rms and phase-lag depend only on three variables: The normalized variations of the seed photon temperature ($\delta T_s = \Delta T_{s}/T_s{}$), the
inner disk temperature ($\delta T_{e} = \Delta T_{e}/T_{e}$) and the phase angle between them ($\phi_{D} = 2\pi f\tau_{D}$, where $f$ is the frequency under consideration). Instead, it is convenient to fit using the three parameters, $\delta T_{e}$, $\tau_{D}$ and the ratio $\delta T_{s}/\delta T_{e}$. Note that these are the same number of parameters as used in the empirical fits described in \S 2.2. Indeed, in the 4\textendash80 keV energy range being considered, the functional form of the fractional rms as predicted by the model is similar to a power-law and the time-lags increase roughly with the logarithm of the energy (Figure \ref{theoryvar2}). Hence it may not be surprising that the reduced $\chi^2$ obtained from the fits of the physical model are similar to the empirical ones (Figure \ref{redchi1}). Thus, the equations (\ref{eq1}) and (\ref{eq2}) defined in \S 2.2 are found to be satisfactory.
\\[6pt]
However, it is important to realize that the  behaviour of the energy dependent rms and time-lags as predicted by the model, depend on the time averaged spectral fit parameters, the inner disk temperature $kT_{e}$ and the photon index $\Gamma$. Here we discuss the sensitivity of the model fitting to these spectral parameters. As an example,
for the August data, $kT_{e} \sim 1.06$ keV and $\Gamma \sim 1.46$ from spectral fitting and the reduced $\chi^2$ obtained for the temporal fitting at 1 Hz is $\sim 1.5$. We find that the reduced $\chi^2$ is insensitive to lower values of $kT_{e}$ and increases sharply for values greater than $\sim 1.4$ keV.
Similarly, the reduced $\chi^2$ is insensitive to the precise value of $\Gamma$, but rises steeply if $\Gamma$ is taken to be larger than 2. While these results are qualitatively same for all observations, there is a difference for the January data. The best fit spectral parameters are $kT_{e} \sim 0.46$ keV and $\Gamma \sim 1.56$ and the temporal fitting is
insensitive to these values except if $kT_{e} > 0.8$ keV, where the reduced $\chi^2$ becomes larger than 2. On the other hand, the temporal fitting is insensitive to large values of $\Gamma$ and the reduced $\chi^2$ becomes larger than 2, for $\Gamma < 1.4$. It is interesting to note that the temporal modelling requires a smaller $kT_{e} \sim 0.5$ keV as compared to the other data sets ($kT_{e} \sim 1.0$ keV) and this is borne out by the spectral fitting. Note that this was possible only due to low energy coverage provided by SXT on-board {\it AstroSat}. In other words for the January data, based on the temporal behaviour as recorded by LAXPC, the model requires a lower inner disk temperature, which is confirmed by the SXT spectral analysis.
\\[6pt]
The results of the model fitting are presented in Figure \ref{model_var_fig}, where $\delta T_{e}$,
$\tau_{D}$ and the ratio $\delta T_{s}/\delta T_{e}$  are plotted as a
function of frequency. The variation of $\delta T_e$ shows two peaks for January data and three peaks for the rest, which resembles the features exhibited by their respective PDS.
The time delay ($\tau_{D}$) between the variation of the disk temperature and that of the Comptonizing cloud is a function of frequency. The ratio $\delta T_{s}/\delta T_{e}$ represents the attenuation of the propagation from the disk to the corona. The ratio is roughly flat and falls slightly at higher frequencies.
\\[6pt] 
The magnitude of $\delta T_{e}$ for the January data is less compared to the other observations, especially for low frequencies. On the contrary, the ratio $\delta
T_{s}/\delta T_{e}$ is significantly higher at low frequencies as compared to the other observations. Thus the decrease in  $\delta T_{e}$ seems to be due to higher attenuation of the cold disk variation as it propagates to the inner hot disk. Also, $\tau_{D}$ between the variations is larger for the January data as compared to the rest. 
\\[6pt] 
All these suggest a possible scenario where for the January observation the disk is truncated at a larger radius than for the others giving rise to (i) a lower inner disk temperature ($~$0.5 rather than $~$1
keV), (ii) higher attenuation as reflected in a larger $\delta T_{s}/\delta T_{e}$ and (iii) a larger time-lag between the cold disk variation and the hot inner flow. This is different from earlier interpretation \citep[e.g.][]{Axe05} where for spectral states similar to the January observations it has been argued that the disk truncates at a smaller radii because the PDS extends to higher frequencies (Figure \ref{powspec}). However, we consider the above three reasons to be a stronger indication that the inner radius is truncated at a larger radius.
\subsection{Comparison with other propagation models}
The basic idea of stochastic fluctuations propagating from outer radii to the inner region has been the primary model to explain the energy dependent continuum variability (as opposed to the quasi-periodic variability) of X-ray binaries \citep{Rap17, Mah18}. The primary difference between the model presented here and the earlier works \citep{Sha76, Hon96}, is that here we consider the Comptonized spectrum to arise from a single region (i.e. one zone), while in the others an emissivity profile in the radius had to be assumed. The energy dependent time-lags arise because the radial emissivity profile is assumed to be different for low and high energy photons. In the model presented here, we say that they arise due to the intrinsic time-lag between the seed photon fluctuations and the subsequent variation of the heating rate of the inner disk. In this section, we discuss the strengths and limitations of these two approaches.
\\[6pt]
The primary spectral component for the hard state is thermal Comptonization by a high temperature medium having a scattering optical depth of the order unity. A typical photon scatters several times and since the optical depth is small, the scattering sites would be located all over the region and not localized. This suggests heuristically that the spectrum arises from all over the region. Moreover, such a hot region would perhaps be isothermal \citep{Sha76}, especially if conductivity is taken into account \citep{Hon96}. Hence it is not clear whether a strong energy dependent emissivity profile would arise in such a medium and such assumptions maybe considered ad-hoc till detailed simulations including the radiative process shows that would happen. 
\\[6pt]
Thus the assumption of single zone region for the spectrum seems to be a better justifiable simpler assumption. However, the price one pays for this is that now the frequency dependent time-lags have to be intrinsic to the propagation model. If one assumes that there is an energy dependent emissivity profile then the frequency and energy dependence of the time-lag between energy bands arise because the different frequencies arise from different regions of the disk and hence the propagation time between them changes. On the other hand, in the model presented here, it has to be assumed that the propagation delay between the truncation radius and the inner region is intrinsically frequency dependent. However, frequency dependent propagation times are expected in cylindrical geometry for simple
sound wave \citep{Mis00}. Even for the standard viscous diffusion process in accretion flow this seems to hold true as revealed by numerical simulations. 
\\[6pt]
According to \citet{Cow14}, the time-lag expected in stochastic propagation is frequency dependent. \citet{Ahm18} have shown that for standard stable gas pressure dominated disk the propagation time-scale is frequency dependent for frequencies higher than the local viscous time-scale. While these results are indicative, one should bear in mind that they are for standard optically thick disks, and it is not certain how quantitatively different would the results be for a hot optically thin, geometrically thick flow and when the propagation time being considered is from the transition region on wards. Nevertheless, there seems to be evidence that the propagation time is frequency dependent and the observed frequency dependence of the time delay between energy bands may be caused by this intrinsic dependence as assumed in this work. Clearly, more detailed numerical simulation of the temporal response
of a truncated disk with a hot inner flow, should be able to justify these
assumptions further. 
\\[6pt]
Another difference between the present model and earlier ones is that in earlier works \citep{Rap17, Mah18}, low frequency component seen in the PDS is thought to be generated in the outer cold disk, while
the high frequency one originates in the hot inner flow. This identification is not completely satisfactory since for some hard state observations of Cygnus X-1, the PDS are better represented by three and not two components (\S 2.2) and it is not clear where the third component arises in this spectrum. In the model presented here, all the variability originates at or beyond the transition radius and no explanation for the shape of the PDS is offered. The PDS here may be due to the complex structure of the transition radius causing different frequencies to propagate through at different dampened levels. 
\\[6pt]
Earlier studies have shown that frequency dependent time-lag between two energy bands is not really a power-law but shows step like behaviour coincident with the shape of the PDS as shown in Figure 2 \citep{Kot01, Mis17}. It should be noted that none of the propagation models provide a reasonable explanation for this behaviour, which implies that the model lacks perhaps several important physical effects. In the next section, we highlight some of the features which have been neglected in the model presented here and speculate that some of them may be the key to make a more quantitative comparison with the data.
\section{Discussion}
We report on the analysis of {\it {AstroSat}} observations of Cygnus X-1 in the standard hard state, during January - October 2016. The broad band spectrum (1\textendash80 keV) observed by SXT and LAXPC is well represented by a dominant thermal Comptonization component with reflection  and disk emission. The LAXPC reveals a PDS which can be modelled roughly as having three (or two) components, which is a known characteristic of the system in the hard state. For a range of frequencies, LAXPC provides an unprecedented view of the energy dependent fractional rms and time-lag from 4\textendash80 keV. {\it {AstroSat}}'s capability of measuring the broad band time-averaged spectrum and the energy dependent temporal behaviour of the system allows one to quantitatively fit both the spectral and temporal data. 
\\[6pt]
We introduce a simple stochastic propagation model to quantitatively fit the energy dependent fractional rms and time-lag at different frequencies. The geometry is that of a truncated standard disk and an hot inner flow which produces the primary spectral component by Comptonizing the outer disk photons to higher energies. An oscillation which originates in the outer regions causes the temperature of the truncation radius to vary causing variations in the Comptonized spectrum. Subsequently, after a time-delay, the perturbation reaches the inner region causing a change in the electron temperature of the inner region and hence a variation in the spectrum. We show that this simple model which is characterized  by only three parameters, (the normalized amplitudes of temperature of the truncation, inner radii and the time-lag between them) can quantitatively explain the observed energy dependent rms and time-lag.
\\[6pt]
There are several, perhaps important physical effects which have not been included in this simple picture, which in the very least would make quantitative if not qualitative difference to the results presented here. These are pointed out below with a brief discussion: 
\\[4pt]
\noindent $\bullet$ The contribution of the reflection  and disk components have been neglected in the timing analysis. While the disk contribution
is small in the LAXPC 4-80 keV band, the flux ratio of the reflection component
to the total one varies from 10 to 50\% for different observations. Effect of the reflection component on timing analysis is complex and
future improved analysis should consider this issue.
\\[4pt]
\noindent $\bullet$ The light travel time effects, especially due to multiple scattering of photons in the inner flow have been neglected. For a ten solar mass black hole and if the truncation is located at say 50 Schwarzchild radius, the light crossing time-scale through the region would be around $\sim 5$ milliseconds. Scattering effects may delay the photons relative to each other by a few times this time scale. While this should not be important for low frequencies ($< 1$ Hz) where the time-lag is much larger, this may seriously effect the  high frequency results, unless the size of the inner flow is much smaller than 50 Schwarzchild radius. 
\\[4pt]
\noindent $\bullet$ The linear analysis undertaken here is valid when the amplitudes of oscillations are small. However, the fractional rms variability of the flux is as high as 20\% and hence non-linear effects may be important. 
\\[4pt]
\noindent $\bullet$ In contrast to the earlier works which have assumed an emissivity profile, we assume here that all the Comptonized photons arise from a single homogeneous zone. In reality, both the spatial emissivity profile and the time-lag between the temperatures of the seed photon and Comptonizing plasma, may be contributing to the observed time-lag. \\[4pt]
\noindent $\bullet$ Irradiation of the outer disk by the Comptonized spectrum has been neglected. Variation of the Comptonized spectrum may change the temperature at the truncation radius and this would lead to decrease in the observed time-lags and other effects.
\\[6pt]
Despite these and may be other limitations, it is heartening to see that the simple model is able to quantitatively fit the energy dependent temporal features. The fitting also provides physical measures of the time a perturbation takes to travel from the truncation radius to the inner regions as well as the normalized variation of the temperatures. Future, {\it {AstroSat}}'s observations of Cygnus X-1 and other X-ray binaries will allow such analysis to be undertaken and warrants improvement of the simple model. The result obtained would become a benchmark to test the predictions of time dependent numerical simulations of a truncated accretion disk.
\section*{Acknowledgements}
We would like to thank the anonymous reviewer for the valued comments and suggestions, which greatly helped in improving the quality of the manuscript. We acknowledge the strong support from Indian Space Research Organization (ISRO), Bengaluru in various aspect of Large Area X-ray Proportional counter (LAXPC) instrument building, testing,  and mission 
operation during payload verification phase. We also acknowledge the support of Tata Institute of Fundamental Research (TIFR), Mumbai central workshop during the  design and testing of the payloads. This work has used the data from the LAXPC and  the Soft X-ray Telescope (SXT) developed at TIFR, Mumbai. The  LAXPC POC and SXT POC at TIFR are thanked for verifying and releasing the data via the ISSDC data archive and providing the necessary software tools. We (MSP, RM, SBG $\&$ BSG) acknowledge Centre for Research - Projects, CHRIST (Deemed to be University), Bengaluru for the research funding through MRPDSC-1721. One of the authors (SBG) thanks the Inter-University Centre for Astronomy and Astrophysics, Pune for Visiting Associateship. 
\\[6pt]
This research has made use of data and/or software provided by the High Energy Astrophysics Science Archive Research Center (HEASARC), which is a service of the Astrophysics Science Division at NASA/GSFC and the High Energy Astrophysics Division of the Smithsonian Astrophysical Observatory.












\bsp	
\label{lastpage}
\end{document}